\documentclass[12pt,a4]{iopart}
\usepackage[english]{babel}

\usepackage{iopams}  
\usepackage{amssymb}
\usepackage[numbers,sort&compress]{natbib}
\usepackage[usenames,dvipsnames,svgnames,table]{xcolor}
\usepackage[hyperindex,colorlinks,breaklinks,allcolors=black]{hyperref}

\usepackage{graphicx,sidecap}

\catcode`@=11 
\renewcommand\tableofcontents{%
  \section*{\contentsname}%
  \@starttoc{toc}%
}
\catcode`@=12

\usepackage{color}

\newcommand{\Teff}{T_{\mathrm{eff}}}

\def\newblock{}

\definecolor{ao(english)}{rgb}{0.0, 0.5, 0.0}
\definecolor{applegreen}{rgb}{0.55, 0.71, 0.0}
\definecolor{cadetblue}{rgb}{0.37, 0.62, 0.63}
\definecolor{cadet}{rgb}{0.33, 0.41, 0.47}
\definecolor{byzantine}{rgb}{0.74, 0.2, 0.64}
\definecolor{orange}{rgb}{1.0, 0.5, 0.0}
\definecolor{magenta}{rgb}{1.0, 0.0, 1.0}

\renewcommand{\vector}[1]{\mathbf{#1}}

\newcommand{\Eq}[1]{Eq.~(\ref{eq:#1})}

\newcommand{\Fig}[1]{Fig.~\ref{fig:#1}}
\newcommand{\Figs}[1]{Figs.~\ref{fig:#1}}

\newcommand{\Sect}[1]{Sect.~\ref{sec:#1}}
\newcommand{\Sects}[1]{Sects.~\ref{sec:#1}}
\newcommand{\sect}[1]{\ref{sec:#1}}

\newcommand{\crit}[1]{{#1}_{\mathrm{c}}}

\begin{document}

\title[Prethermalization and universal dynamics in near-integrable quantum systems]
{Prethermalization and universal dynamics in near-integrable quantum systems}

\author{Tim Langen}
\address{JILA, University of Colorado and NIST, Boulder, CO 80309, USA}
\ead{tim.langen@colorado.edu}

\author{Thomas Gasenzer}
\address{Kirchhoff-Institut f\"ur Physik, Ruprecht-Karls-Universit\"at Heidelberg,\newline Im Neuenheimer Feld 227, 69120~Heidelberg, Germany}
\address{ExtreMe Matter Institute EMMI, GSI Helmholtzzentrum f\"ur Schwerionenforschung GmbH, Planckstra\ss e~1, 64291~Darmstadt, Germany}
\ead{t.gasenzer@uni-heidelberg.de}

\author{J\"org Schmiedmayer}
\address{Vienna Center for Quantum Science and Technology, Atominstitut, TU Wien, Stadionallee 2, 1020 Wien, Austria}
\ead{schmiedmayer@atomchip.org}

\vspace{10pt}
\begin{indented}
\item[]12 May 2016
\end{indented}

\begin{abstract}
We review the recent progress in the understanding of the relaxation of isolated near-integrable quantum many-body systems. Focusing on prethermalization and universal dynamics following a quench, we describe the experiments with ultracold atomic gases that illustrate these phenomena and summarize the essential theoretical concepts employed to interpret them. Our discussion highlights the key topics that link the different approaches to this interdisciplinary field, including the generalized Gibbs ensemble, non-thermal fixed points, critical slowing and universal scaling.  Finally, we point to new experimental challenges demonstrating these fundamental features of many-body quantum systems out of equilibrium.
\end{abstract}

%
%
%
\maketitle
%
%
\tableofcontents

\section{Introduction}
\label{sec:Intro}

The relaxation of isolated quantum many-body systems is a fundamental unsolved problem connecting many different fields of physics. 
Examples range from the dynamics of the early universe and quark-gluon plasmas to coherence and transport in condensed matter physics and quantum information.
Consequently \textit{relaxation} and \textit{non-equilibrium dynamics} encompass a wide range of phenomena on vastly different energy, length, and time scales. 

In the last years, rapid progress in the field was triggered by experiments with ultracold quantum gases \cite{Bloch2008a.RevModPhys.80.885}, which allow probing of the fundamental processes and time scales of these phenomena
\cite{Kinoshita2006a,Hofferberth2007a,Trotzky2012a.NatPhys8.325,Gring2011a,Kuhnert2013a,Langen2013b,Geiger2014a,Langen2015b.Science348.207,Sadler2006a,Vengalattore2008a,Kronjager2010.PhysRevLett.105.090402,Nicklas2011a,Nicklas:2015gwa,Cheneau2012a,Lamporesi2013a,Ronzheimer2013a.PRL110.205301,Braun2014a.arXiv1403.7199B,Corman2014a,Chomaz2015a,Navon2015a.Science.347.167N,Langen2015a.annurev-conmatphys-031214-014548}.
These experiments opened up a new path to testing theoretical concepts, revealing their implications for realistic systems.

In the experiments, an initially prepared, trapped gas is suddenly `quenched' out of equilibrium.
This is typically achieved by a parameter change in the trapping potential or the interaction properties.
\emph{Relaxation}, i.e., the evolution to a new \mbox{(quasi-)} stationary state is subsequently probed through temporally and spatially resolved measurements of suitably selected observables.
Central questions that have been studied range from the existence of \emph{non-thermal steady states} and the microscopic dynamics that establish these states to the emergence of aspects of \emph{macroscopic} classical statistical mechanics from the \emph{microscopic} unitary quantum evolution. 
A fundamental problem thereby is how unitary evolution, which conserves the von Neumann entropy $S=-\mathrm{Tr}[\hat\rho\ln\hat\rho]$ of the full quantum state $\hat\rho$ of an isolated system can lead to relaxation and \emph{thermalization}  (the approach of a steady state well described by a Gibbs ensemble). 

In the establishment of \emph{(quasi-)steady states} conser\-vation laws play a key role. 
Consequently, isolated many-body systems (approximately) described by \emph{quantum integrable models} with a large number of conserved quantities are of particular interest.
Ultracold atomic gases are ideally suited to realize and probe such systems. 
In particular, they allow studying different mechanisms that break the integrability. 
This opens the unique possibility to investigate the intricate relationship between  thermalization and integrability in quantum systems. 

\emph{Emergence of macroscopic physics: The generalized Gibbs ensemble.---}
Information theory provides a particularly powerful approach~\cite{Shannon:1949} to studying such steady states. 
As pointed out by Jaynes in the 1950s, once one realizes that ther\-mo\-dynamic entropy and information entropy are the same concept, one may take entropy as the starting point and consider statistical mechanics as a form of statistical inference \cite{Jaynes1957a,Jaynes1957b}.
From this point of view,  the maximum-entropy state takes the central role, being the least biased estimate possible on given information, or the maximally non-committal configuration with regard to missing information \cite{Jaynes1957a}.
The maximum-entropy principle leads directly to the standard thermodynamical ensembles, which are constrained only by  quantities like energy or particle number~\cite{Huang1987a}.
However, as highlighted by the work of Rigol and collaborators \cite{Rigol2007a}, it has recently become clear that the late-time behavior of isolated integrable quantum many-body systems is fundamentally different from that of non-integrable ones \cite{Rigol2008a,Rigol2009a,Polkovnikov2011a,Eisert2015a,Gogolin2015a}. 
Further conservation laws constrain the dynamics of integrable systems and thus, appropriately \emph{generalized Gibbs ensembles} (GGE) are expected on the grounds of the maximum-entropy argument \cite{Jaynes1957a,Jaynes1957b}.
These GGEs were indeed observed in experiment, in the relaxation of a 1D quantum gas \cite{Langen2015b.Science348.207}.
When deriving statistical ensembles from the microscopic physical properties one usually has to complement the basic equations of motion with further physical assumptions, in particular ergodicity.
Jaynes' statistical, information-based approach is, however, independent of such  assumptions \cite{Jaynes1957a} and, hence, it is  applicable not only to the special subclass of equilibrium states but also to dynamical evolution  processes \cite{Jaynes1957b,Girardeau1969a,Girardeau1970a,Rigol2007a,Rigol2008a}.
The questions arising in the above context have been quickly taken up by experimenters \cite{Langen2015a.annurev-conmatphys-031214-014548}, as will be discussed in detail in \Sect{experiments-near-integrable}.

\begin{figure}
\centering
\includegraphics[width=.52\textwidth]{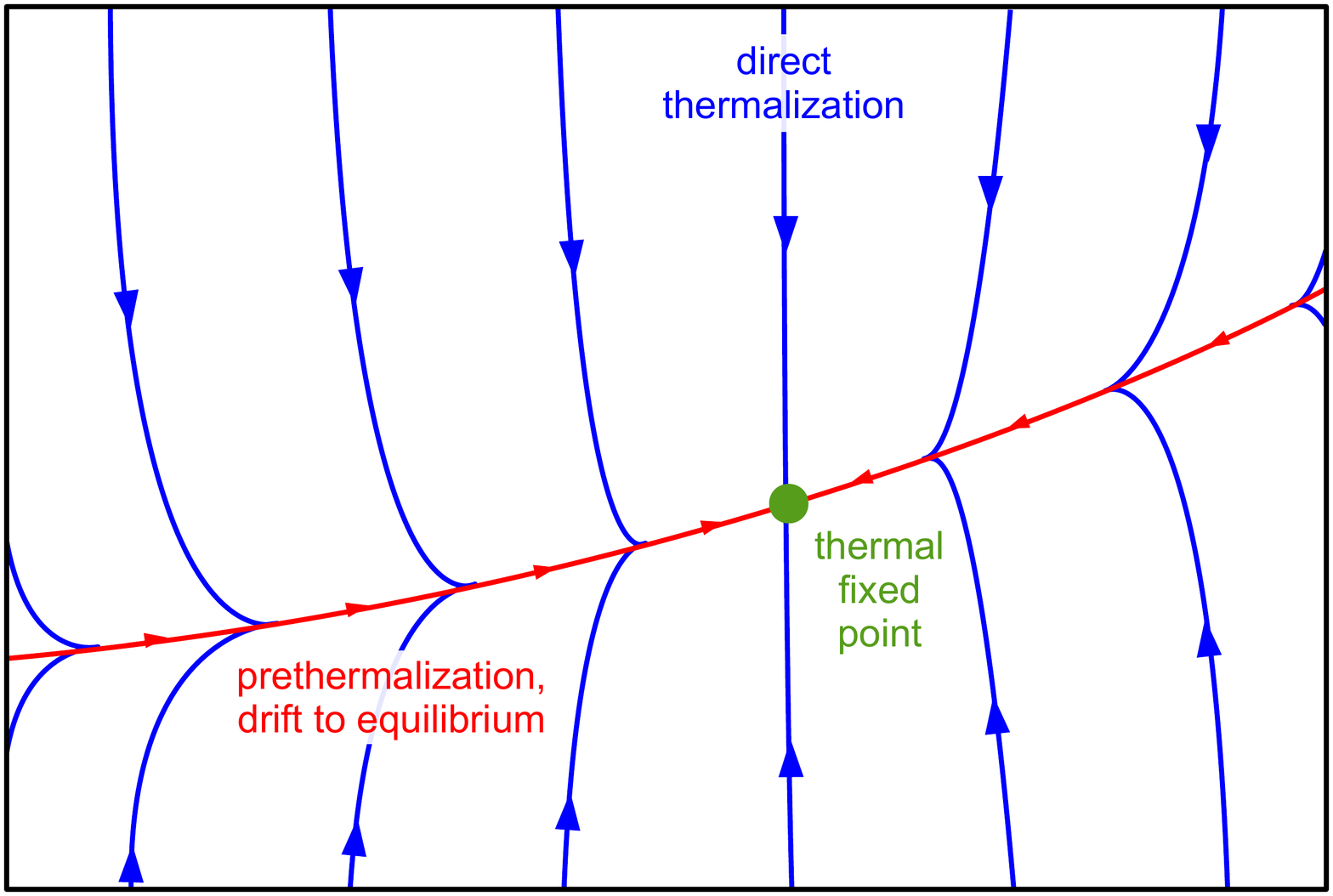} 
\includegraphics[width=.46\textwidth]{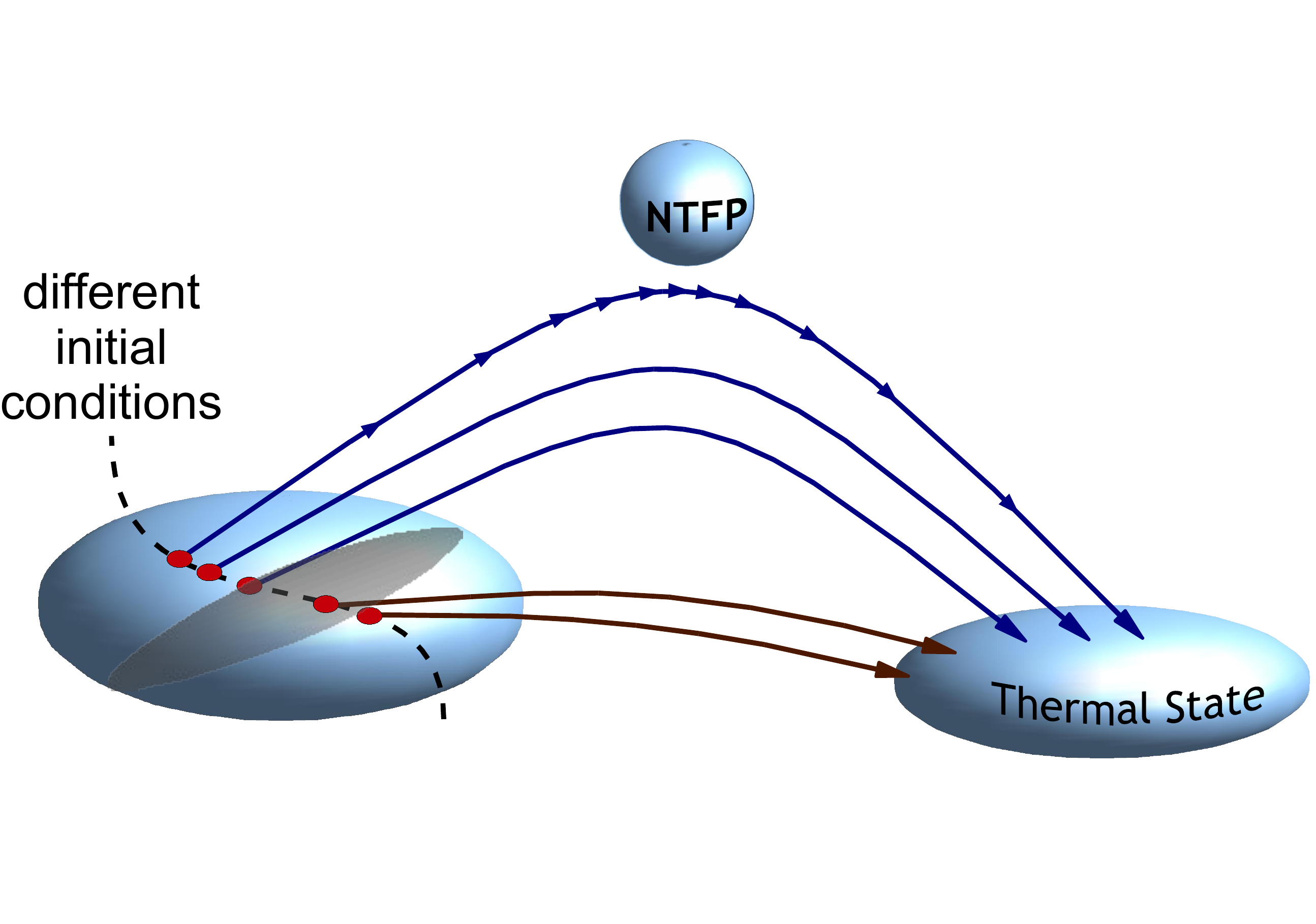} 
\caption{Schematics of prethermalization \cite{Aarts2000a,Berges:2004ce} (left panel, adapted from \cite{Wetterich2012a.privcomm}) and a non-thermal fixed point \cite{Berges:2008wm} (NTFP, right panel, adapted from \cite{Nowak:2012gd}), based on the ideas of a renormalization group flow.
The left panel depicts possible time evolutions in a two-dimensional projection or sub-manifold of the space of many-body states.
The arbitrarily chosen axes represent running `coupling' parameters.
These are related, e.g., to the various Lagrange parameters of a generalized Gibbs ensemble (GGE, cf.~\Sects{gge} and \sect{GGE-Expt}), of which all but the Gibbs-ensemble parameters vanish in the thermal state (green `thermal fixed point'). 
Hence, the space of such trajectories is in general not restricted to the two dimensions chosen in the sketch. 
A near-integrable system quickly approaches a prethermalized state (red line), which, owing to conservation laws, retains memory of the initial conditions and then slowly drifts to the thermal fixed point. 
Depending on the particular choice of initial condition (e.g.~quench), different, non-universal early-time evolutions (blue trajectories) occur before the system starts to show universal behavior.
In contrast, in non-integrable systems, direct thermalization on a single time scale as in kinetic damping is the generic pathway. 
The red drift line can be regarded as a partial fixed point which is approached quickly from generic initial conditions and where the system experiences critical slowing down. 
Along the red line, correlation functions $C(p,t)$ can show scaling behavior in space and time, e.g., $C(p,t)=t^{\alpha}f(t^{\beta}p)$, with a universal scaling function $f$ (cf.~also \protect\Fig{NTFP}).
In the case of prethermalization to a GGE in a (near-)integrable system, one expects $\alpha=\beta=0$ $(\simeq0)$ such that the system gets (almost) stuck when reaching the red line.
Instead of ending at the stable thermal fixed point, the red line could also lead to a partially stable NTFP.
In this case,  the blue `direct thermalization' lines are replaced by trajectories leading away from the NTFP, e.g., towards a different final thermal fixed point (see also sketch in the right panel).
In this situation, more general scaling with $\alpha$, $\beta\not=0$ is expected on the red line, where the algebraic time evolution nonetheless means slowing down of the evolution (cf. arrows in right panel). 
}
\label{fig:relaxation}
\end{figure}

\emph{Microscopic relaxation processes: Prethermalization.---}
Given the new types of stationary states discussed above, it is also important to better understand the \emph{relevant micro\-scopic processes} leading to these states. However, a general framework for the relaxation dynamics of isolated quantum many-body systems quenched far out of equilibrium largely remains an open problem.

\newpage
The key question is which types of non-trivial `pathways' such systems can, in general, take under the influence of conservation laws.
In particular for integrable and near-integrable quantum systems, one expects special relaxation characteristics.
It is then especially interesting  to find those degrees of freedom which need to be taken into account when performing a maximum-entropy analysis of the relaxed states.
Generically, relaxation within a simple kinetic framework is expected to be described by exponential laws in time, with a rate defining the time scale.
In contrast, (near-) integrable systems relax in a way that is characterized by at least  two time scales.
This is due to  a distinction between fast and slow processes, such that  `prethermalization' \cite{Aarts2000a,Berges:2004ce} occurs as a prelude to asymptotic relaxation to  a thermal or any other type of equilibrium state.

As will be discussed in more detail in \Sect{pretherm}, prethermalization, in most cases studied so far, has been associated with a fast relaxation to a metastable state in which much of the phase coherence of the initial state is lost.
Often, and in particular for the experiments described in \Sect{SplitRelax-Expt}, this process can be traced back to mean-field dephasing of weakly interacting quasi-particles.
In the prethermalized state, bulk quantities such as the total kinetic energy can already be close to their final equilibrium values while other observables are not.
The latter include, e.g., quasi-particle occupation numbers fixed by initial conditions, giving rise to the description of the prethermalized state in terms of a GGE.
For the special case of integrable systems for which it is possible to identify a description in terms of a free field theory, the dephasing of these fields leads directly to a final steady state.
This steady state is the limit of the prethermalized state for vanishing quasiparticle interactions.

\emph{Universal (scaling) dynamics and non-thermal fixed points.---}
Prethermalization has first been proposed on the basis of ideas of a renormalization-group flow \cite{Aarts2000a,Berges:2004ce,Bettencourt:1997nf,Bonini1999a}, see \Fig{relaxation} for an illustration.
In the wider con\-text of relaxation dynamics, `prethermalization' means the approach of any (partially) universal intermediate state which is still out of equilibrium with respect to asymp\-totically long time evolution.
The state being \emph{universal} means that it can be described mathematically in terms of a limited set of parameters and/or functions which only depend on a corresponding set of symmetries obeyed in the dynamical evolution of the state.
These characteristics are, however, independent of the particular physical realization and of the specific initial configuration resulting from the quench.
If the state is only \emph{partially universal}  the dynamics is dominated by the universal characteristics while  non-universal properties remain depending, e.g., on the particular initial state (also the final thermal state being partially universal in this respect). 

As an example, during the prethermalization stage, measurable correlations could be well reproduced by assuming the system to be described by a GGE containing a limited set of conserved operators.
These operators are directly related to intrinsic symmetry properties of the system's Hamiltonian, while, in contrast, the values of the Lagrange multipliers pertaining to the conserved quantities are non-universal as they depend on values of these quantities in the initial state.

Another example is prethermalization to an algebraically-slowly evolving state.
For instance, phase-ordering kinetics arising after a quench across a symmetry-breaking phase transition could be described in terms of universal distribution functions showing self-similar coarsening evolution in space and time \cite{Bray1994a}.
The same is true for wave-turbulent kinetics where the self-similar evolution appears in the form of cascading transport of e.g.~energy between different scales \cite{Zakharov1992a}, similar to classical fluid turbulence.

Besides slowing-down, the property common to these evolutions is scaling behavior, with evolution time as scaling parameter.
This scaling is reminiscent of equilibrium criticality at a continuous phase transition \cite{Goldenfeld1992a,ZinnJustin2004a,Ma2000a.ModernTheoryCritPhen}:
the time evolving system on average rescales in space with a power of the time parameter, which looks like zooming in or out the field of view of a microscope in real time.
Such scaling and slowing are, to a certain extent, analogues of the universality in equilibrium critical phenomena  in nonequilibrium systems \cite{Hohenberg1977a,Bray1994a,Sachdev2000a,Henkel2008a.NonEqPhaseTransitions,Taeuber2014a.CriticalDynamics}.  
Recent progress has been made in the context of non-thermal fixed points \cite{Berges:2008wm,Berges:2012ty,Nowak:2013juc}, see \Sect{universal} below.  
The possibility of categorising systems into generalized `universality classes' associated with possible new critical exponents is a fascinating prospect and the subject of current research.

In the following we discuss in more detail the recent developments in the conceptual and experimental  studies of prethermalization, GGEs, and universal dynamics in (near-) integrable quantum-gas systems. 
In \Sect{pretherm} we extend our summary of the theoretical status, discussing in particular the connection between prethermalization  and universal dynamics. Experiments on relaxation dynamics in near-integrable systems, prethermalization and GGEs are the subject of \Sect{experiments-near-integrable}. We close with an outlook on developments under way.

\section{Prethermalization, non-thermal fixed points and universal dynamics}
\label{sec:pretherm}
We start by reviewing the theoretical basis for discussing relaxation of quantum systems described by (near-)integrable models. 
Important concepts are the GGE resulting from a pure statistical approach, as well as, on the microscopic dynamical level, prethermalization, the approach of non-thermal fixed points, and universal (scaling) dynamics.  
See also other contributions to this volume \cite{Ilievski:2016fdy,
Calabrese:2016xau,
Cazalilla2016a.160304252C,
Caux2016a.160304689C,
Essler2016a,
Vasseur2016a.160306618V,
DeLuca2016a.160308628D,
Vidmar2016a.160403990V} for more detailed discussions of the theoretical background.

\subsection{Statistical description of the stationary state}
\label{sec:statistical}

\subsubsection{(Near-)integrable quantum systems}
\label{sec:integrable}
Integrability and its consequences have been crucial for understanding the process that inhibit or allow thermalization in classical mechanics. 
In one of the first numerical experiments, Fermi, Pasta, and Ulam studied the evolution of a chain of harmonic oscillators with non-linear couplings thereby observing quasi-periodic instead of ergodic behavior. 
This surprising result can be ascribed to the \emph{integrability} of the model, which allows for quasi-periodic motion instead of thermalization \cite{Berman2005a}. 
Integrability of classical model systems results from the existence of a full set of conserved quantities, which restrict the evolving system to a toroidal sub-region of the total phase space. 
The above findings have ultimately led to the development of chaos theory, which became the basis for the understanding of classical thermalization \cite{Rigol2010b.arXiv1008.1930}. 

In quantum systems the meaning of integrability is less clear \cite{Caux2011a.JSMTEP02023}. 
Different definitions that have been used include the existence of a complete set of algebraically independent mutually commuting conserved operators  \cite{VonNeumann:791332} (with the flaw that it is, according to the spectral theorem, fulfilled by any hermitian Hamiltonian as one could simply take the projectors on the energy eigenstates \cite{Caux2011a.JSMTEP02023}).
Other definitions are based on the equivalence of the system to a set of independent harmonic oscillators, on the occurrence of non-diffractive scattering processes \cite{Sutherland1998a}, or on the existence of a Bethe-Ansatz or otherwise known exact solution \cite{Korepin1997a,Mussardo2009a}. 
The latter criterion of `integrability' does not necessarily have intrinsic validity but, nevertheless, usually implies the properties required by integrable systems.
Examples of quantum integrable models under these criteria are the Lieb-Liniger model of $N$ point-interacting bosons on a line \cite{Lieb1963a,Lieb1963b}, the one-dimensional Hubbard model \cite{Shastry1986a},  and variants of the Heisenberg model \cite{Korepin1997a}.
Note, however, that the above criteria are not one-to-one identical implying, e.g., that non-integrable models can nonetheless be analytically solvable \cite{Braak2011a.PhysRevLett.107.100401}.
As will be discussed below, in the context of prethermalization, it is the number of spatially (quasi-)local conserved charges which appears to matter.

The question whether and how quantum systems thermalize has regained much impetus through ultracold atomic gases described by (near-)integrable mathematical models.
A first explicit demonstration of a strongly interacting one-dimensional Bose gas not showing any thermalization over a period corresponding to many collision times was delivered in the experiment by Kinoshita et al.~\cite{Kinoshita2006a}.
Their system is in principle described by the integrable Lieb-Liniger (LL) model \cite{Lieb1963a,Lieb1963b}.
In the specific experimental environment, it is, however, expected to become weakly non-integrable due to the presence of the transverse degrees of freedom and the additional longitudinal harmonic trapping potential. 
Details of this experiment are discussed in Sect.~\ref{sec:experiments-near-integrable}.
Previous studies had demonstrated the applicability of a description in terms of the hard-core or Tonks-Girardeau limit \cite{Girardeau1960a} of the LL model \cite{Paredes2004a,Kinoshita2004a}.
In this limit, the model maps to a system of non-interacting spin-less fermions, which is analytically solvable and due to the absence of scattering does not thermalize.
These experiments sparked a new research direction, asking whether the time evolution under the constraint of conserved quantities could at all lead to a thermalized final state \cite{Rigol2007a,Rigol2008a,Rigol2009a}.

\subsubsection{The generalized Gibbs ensemble }
\label{sec:gge}

Despite following a strictly unitary evolution, generic isolated, non-integrable quantum systems are now expected to exhibit thermalization in accordance with classical experience.
If only the total energy is conserved, one observes chaotic behavior and thermalization, i.e., local observables relax to a canonical or Gibbs ensemble with an effective temperature \cite{Deutsch1991a.PhysRevA.43.2046,Srednicki1994a.PhysRevE.50.888,Tasaki1998a.PhysRevLett.80.1373,Berges:2000ur,Berges:2001fi,Berges:2002wr,Popescu2005a,Calabrese2005a.JSMTE.P04010,Gasenzer:2005ze,Eisert2006a.PhysRevLett.97.150404,Berges:2007ym,Rigol2007a,Calabrese2006a.PhysRevLett.96.136801,Calabrese2007a,Rigol2008a,Rigol2009a,Rigol2010a.PhysRevA.82.011604,Biroli2010a.PhysRevLett.105.250401,Banuls2011a.PhysRevLett.106.050405,Eckstein2009a.PhysRevLett.103.056403,Rigol2014a.PhysRevLett.112.170601,Cardy:2015xaa,Polkovnikov2011a,Eisert2015a,Gogolin2015a}. 
A common picture thereby is that a finite region of a system is thermalized by interaction with the rest acting as a bath.
The question, however, why a non-integrable quantum system, in contrast to integrable ones, can be successfully described by such a small number of conserved quantities is a still a matter of intense research.

Despite the lack of a unique definition of integrability, it is now generally accepted that in the quantum case, as in the classical, additional conserved quantities may slow down if not even inhibit thermalization. 
However, even in the absence of strict thermalization, relaxation and the emergence of thermal-like properties are still possible. 
Reviving the statistical arguments by Jaynes \cite{Jaynes1957a,Jaynes1957b}, Rigol et al.~demonstrated, for the exactly solvable case of hard-core bosons on a lattice, that relaxation to an equilibrium state occurs.
This equilibrium state carries more memory of the initial conditions than simply energy and particle number \cite{Rigol2007a,Rigol2008a,Rigol2009a}.
They conjectured that the  presence of non-trivial conserved quantities puts constraints on the available phase space of a system, which also strongly affect the dynamics:
Consider an initial state $|\Psi(0)\rangle$ of a translationally invariant system described by an integrable model with Hamiltonian $\hat H$.
Then, stationary $n$-point correlation functions of local operators $\hat\mathcal{O}_{a}(x)$ in the thermodynamic limit are given by a generalized Gibbs ensemble (GGE),
\begin{equation}
\lim_{t\to\infty}
\langle\Psi(t)|\prod_{a=1}^{n}\hat\mathcal{O}_{a}(x_{a})|\Psi(t)\rangle
= \Tr\Bigg[\hat\rho_{\mathrm{GGE}}\prod_{a=1}^{n}\hat\mathcal{O}_{a}(x_{a})\Bigg],
\label{eq:ggecorr}
\end{equation}
with $|\Psi(t)\rangle=\exp(-i\hat Ht/\hbar)|\Psi(0)\rangle$ and
\begin{equation}
\hat \rho_{\mathrm{GGE}} = \frac{1}{Z}\exp\Big(-\sum_m\lambda_m\,\hat{\mathcal{I}}_m\Big).
\label{eq:gge}
\end{equation}
Here $\hat{\mathcal{I}}_m$ denotes a full set of conserved quantities, $Z=\mathrm{Tr}\exp(-\sum_m\lambda_m\hat{\mathcal{I}}_m)$ is the partition function and $m$ is a positive integer. 
A separate Lagrange multiplier $\lambda_{m}$ is associated with each of the conserved quantities. 
These numbers are obtained by maximisation of the von Neumann entropy $S=-\mathrm{Tr}[\hat\rho_{\mathrm{GGE}}\ln\hat\rho_{\mathrm{GGE}}]$, under the condition that the expectation values of the conserved quantities are fixed to their initial values in the thermodynamic limit, 
\begin{equation}
\mathrm{Tr}[\,\hat{\mathcal{I}}_m\hat\rho_{\mathrm{GGE}}\,] = \langle\hat{\mathcal{I}}_m\rangle(t=0).
\label{eq:gge2}
\end{equation}
The emergence of such a maximum-entropy state is not in contradiction to a unitary evolution according to quantum mechanics \cite{Jaynes1957b}.
It rather reflects that the true quantum state is indistinguishable from the maximum-entropy ensemble with respect to a set of observables~\cite{Polkovnikov2011a}.
It was pointed out in \cite{Barthel2008a.PhysRevLett.100.100601,Gogolin2011a} that for the GGE conjecture to hold it is essential to restrict the observables to finite subsystems.
Hence, the GGE describes the long-time limit of sufficiently \emph{local} observables.

The GGE is a direct generalization of the well-known thermodynamical ensembles.
In cases where only the total energy and the particle number are conserved, it reduces to the grand-canonical ensemble, where temperature ($\lambda_{1}=\beta=1/k_{B}T$) and chemical potential ($\lambda_{2}=-\beta\mu$) play the role of the respective Lagrange multipliers~\cite{Huang1987a}. 
We emphasize that in the cases described here, in most of the basic examples, there is only one species of undistinguishable particles such that there is no natural number conservation arising from intrinsic particle properties.
The conserved quantities can, e.g., measure functions of the fractions of particles in different collective motional states (e.g.~quasiparticle momentum modes).
More generally, they can also be given by correlations of higher order in some fundamental field describing the system's properties.

A question not yet fully resolved is precisely which conserved quantities $\hat{\mathcal{I}}_m$ need to be included in the definition of the GGE.
The current belief is that they should be constructed out of local or quasi-local conserved operators \cite{Essler2016a,Ilievski:2016fdy}, i.e.~$\hat{\mathcal{I}}_m=\sum_{j}i_{m}(j)$, where the sum runs over, e.g., the sites of a lattice, and $i_{m}(j)$ acts non-trivially only around the site $j$ if the charge is local, while it has exponential tails if it is quasi-local.

Consider the case of a free theory, in which Fock mode operators $\hat a_{p}$, $\hat a_{p}^{\dagger}$ diagonalize  the Hamiltonian, $\hat H = \sum_{p}\omega_{p}\hat a_{p}^{\dagger}\hat a_{p}$, with mode frequencies $\omega_{p}$.
Conserved quantities may then be written in the additive form $\hat{\mathcal{I}}_m=\sum_{p}f^{(m)}_{p}\hat a_{p}^{\dagger}\hat a_{p}$ with a set of scalar functions $f^{(m)}_{p}$.
(Note that in free models, the above local charges are linearly related to the mode occupation operators $\hat a_{p}^{\dagger}\hat a_{p}$ such that either of them can be used to define the GGE, with \Eq{gge2} applying to the densities.)
While total energy and particle number are obtained with $f_{p}=\omega_{p}$ and $f_{p}=1$, respectively, further conserved quantities of the free theory are, e.g., simply the occupation numbers themselves, $f_{p}^{(m)}=\delta_{pm}$.
In this case, time evolution is identical to dephasing, leading to stationary values of operators on finite subsystems.

In the case of interacting theories, however, the question of relevant conserved quantities is not yet fully settled  \cite{Bertini2014a.JSMTE.P10035}.
The question which are the relevant conserved quantities has been discussed intensely \cite{Fagotti2013b.1742-5468-2013-07-P07012,DeNardis2014a.PhysRevA.89.033601,Fagotti2014a.PhysRevB.89.125101,Wouters2014a.PhysRevLett.113.117202,Pozsgay2014a.PhysRevLett.113.117203,Goldstein2014a.PhysRevA.90.043625}, in particular the question whether or not quasi-local  conserved operators are required in the construction of GGEs \cite{Fagotti2013a.PhysRevB.87.245107,Ilievski2015a.PhysRevLett.115.157201,Essler2015a.PhysRevA.91.051602,Essler2016a,Ilievski:2016fdy}.

The GGE is now believed to be the final state of relaxation for generic quantum integrable systems 
\cite{Polkovnikov2011a,
Eisert2015a,
Gogolin2015a,
Girardeau1969a,Girardeau1970a,
Rigol2006a,
Cazalilla2006a,
Rigol2007a,
Calabrese2007a,
Barthel2008a.PhysRevLett.100.100601,
Eckstein2008a,
Gangardt2008a,
Iucci2009a,
Fioretto2010a,
Mossel2010a,
Gogolin2011a,
Pozsgay2011a,
Calabrese2011a.PhysRevLett.106.227203,
Calabrese2012a,
Calabrese2012b,
Cazalilla2012a.PhysRevE.85.011133,
Mossel2012a,
Caux2012a.PhysRevLett.109.175301,
Essler2012a.PhysRevLett.109.247206,
Gramsch2012a.PhysRevA.86.053615,
Caux2013a.PhysRevLett.110.257203,
Fagotti2013a.PhysRevB.87.245107,
Collura2013a.PhysRevLett.110.245301,
Mussardo2013a.PhysRevLett.111.100401,
Pozsgay2013a,
Fagotti2013b.1742-5468-2013-07-P07012,
Fagotti2014a.PhysRevB.89.125101,
Cardy2014a.PhysRevLett.112.220401,
Mierzejewski2014a.PhysRevLett.113.020602,
Wouters2014a.PhysRevLett.113.117202,
Pozsgay2014a.PhysRevLett.113.117203,
DeNardis2014a.PhysRevA.89.033601,
Kormos2014a.PhysRevA.89.013609,
Sotiriadis2014a.1742-5468-2014-7-P07024,
Goldstein2014a.PhysRevA.90.043625,
Zill2015a.PhysRevA.91.023611,
Ilievski2015a.PhysRevLett.115.157201,
Essler2015a.PhysRevA.91.051602,
Cardy:2015xaa,
PerarnauLlobet2015a.arXiv151203823P,
Kastner2015a,
Zill:2016boi,
Ilievski:2016fdy,
Calabrese:2016xau,
Cazalilla2016a.160304252C,
Caux2016a.160304689C,
Essler2016a,
Vidmar2016a.160403990V}. 
These studies focused on many different models and theories, including the 
Luttinger model 
\cite{Cazalilla2006a,
Iucci2009a,
Cazalilla2012a.PhysRevE.85.011133,
Cazalilla2016a.160304252C}, 
hard-core bosons in one dimension 
\cite{Rigol2006a,
Rigol2007a,
Rigol2009a,
Gangardt2008a,
Gramsch2012a.PhysRevA.86.053615,
Kormos2014a.PhysRevA.89.013609}, 
conformal field theories 
\cite{Calabrese2007a,
Cardy:2015xaa,
Calabrese:2016xau},
the infinite-dimensional Falicov-Kimball model 
\cite{Eckstein2008a}, 
the Lieb-Liniger model 
\cite{Mossel2012a,
Caux2012a.PhysRevLett.109.175301,
Caux2013a.PhysRevLett.110.257203,
Collura2013a.PhysRevLett.110.245301,
DeNardis2014a.PhysRevA.89.033601,
Sotiriadis2014a.1742-5468-2014-7-P07024,
Goldstein2014a.PhysRevA.90.043625,
Zill2015a.PhysRevA.91.023611,
Zill:2016boi,
Caux2016a.160304689C}, 
the quantum Ising chain in a transverse field 
\cite{Calabrese2007a,
Calabrese2011a.PhysRevLett.106.227203,
Calabrese2012a,
Calabrese2012b,
Cazalilla2012a.PhysRevE.85.011133,
Essler2012a.PhysRevLett.109.247206,
Fagotti2013a.PhysRevB.87.245107,
Kastner2015a,
Essler2016a}, 
and Heisenberg spin chains 
\cite{Mossel2010a,
Gogolin2011a,
Cazalilla2012a.PhysRevE.85.011133,
Pozsgay2013a,
Fagotti2013b.1742-5468-2013-07-P07012,
Fagotti2014a.PhysRevB.89.125101,
Mierzejewski2014a.PhysRevLett.113.020602,
Wouters2014a.PhysRevLett.113.117202,
Pozsgay2014a.PhysRevLett.113.117203,
Goldstein2014a.PhysRevA.90.043625,
Ilievski2015a.PhysRevLett.115.157201,
Ilievski:2016fdy}.
We note that within the conditions set by integrability, various effective `temperatures' have been proposed on the basis of 
static \cite{Calabrese2006a.PhysRevLett.96.136801,
Calabrese2007a,
Rossini2009a,
Mitra2011a.PhysRevLett.107.150602} and 
dynamic \cite{Foini2011a.PhysRevB.84.212404,
Foini2012a.JStatMTE.P09011} properties.
These definitions, however, in general depend on the details of the initial conditions.

We note that ergodicity may also be broken due to many-body localization \cite{Pal2010a.PhysRevB.82.174411,Serbyn2013a.PhysRevLett.111.127201,Vosk2013a.PhysRevLett.110.067204}.
Hence, generalising the concept and taking into account the influence of the external potential, the GGE has also been suggested as a description for such many-body localized states~\cite{Vosk2013a.PhysRevLett.110.067204,Vasseur2016a.160306618V}. 

As will be discussed in detail in \Sect{experiments-near-integrable}, dynamics, relaxation and prethermalization of near-integrable quantum systems have been studied extensively in experiment~\cite{Kinoshita2006a,Trotzky2012a.NatPhys8.325,Gring2011a,AduSmith2013a,Langen2013a.EPJST.217,Geiger2014a,Ronzheimer2013a.PRL110.205301,Langen2015b.Science348.207}. 
These experiments immediately lead to the  important question what happened if certain quantities were only approximately conserved. 
It has been conjectured that in this case an isolated system will first relax to a metastable state described by a GGE~\cite{Moeckel2008a,Moeckel2009aAnPhy.324.2146M,Moeckel2010a.NJP.12.055016,Rosch2008a.PhysRevLett.101.265301,Kollar2011a,Marino2012a.PhysRevB.86.060408,VandenWorm2013a,Marcuzzi2013a.PhysRevLett.111.197203,Essler2014a.PhysRevB.89.165104,Nessi2014a.PhysRevLett.113.210402,Fagotti2014b,Brandino2015a.PhysRevX.5.041043,Bertini2015b,Babadi2015a.PhysRevX.5.041005,Bertini2015a.PhysRevLett.115.180601,Buchhold2015a.arXiv151003447B}. 
In the context of near-integrable systems, this behavior has been related to prethermalization (cf.~\Sect{prethermalization} below; it can also be regarded as an alternative definition of prethermalization).
Eventual thermalization happens, in contrast, on a much longer time scale \cite{Stark2013a.1308.1610}. 
It thereby remains unclear in which way and how far integrability has to be perturbed to allow for actual thermalization \cite{Bertini2015a.PhysRevLett.115.180601}.
This question is of strong conceptual interest, in particular as the related problem has been very well studied in classical mechanics. 
In that case, the famous Kolmogorov-Arnold-Moser (KAM) theorem quantifies the effect of a weak non-integrability on the dynamics \cite{Polkovnikov2011a}. 
No corresponding theorem is known in the quantum case \cite{Brandino2015a.PhysRevX.5.041043}.
\begin{figure}[t]
\flushleft{\small\quad(a)\hspace{0.47\textwidth}(c)}\\[-4.5ex]
\centering
\includegraphics[width=.46\textwidth]{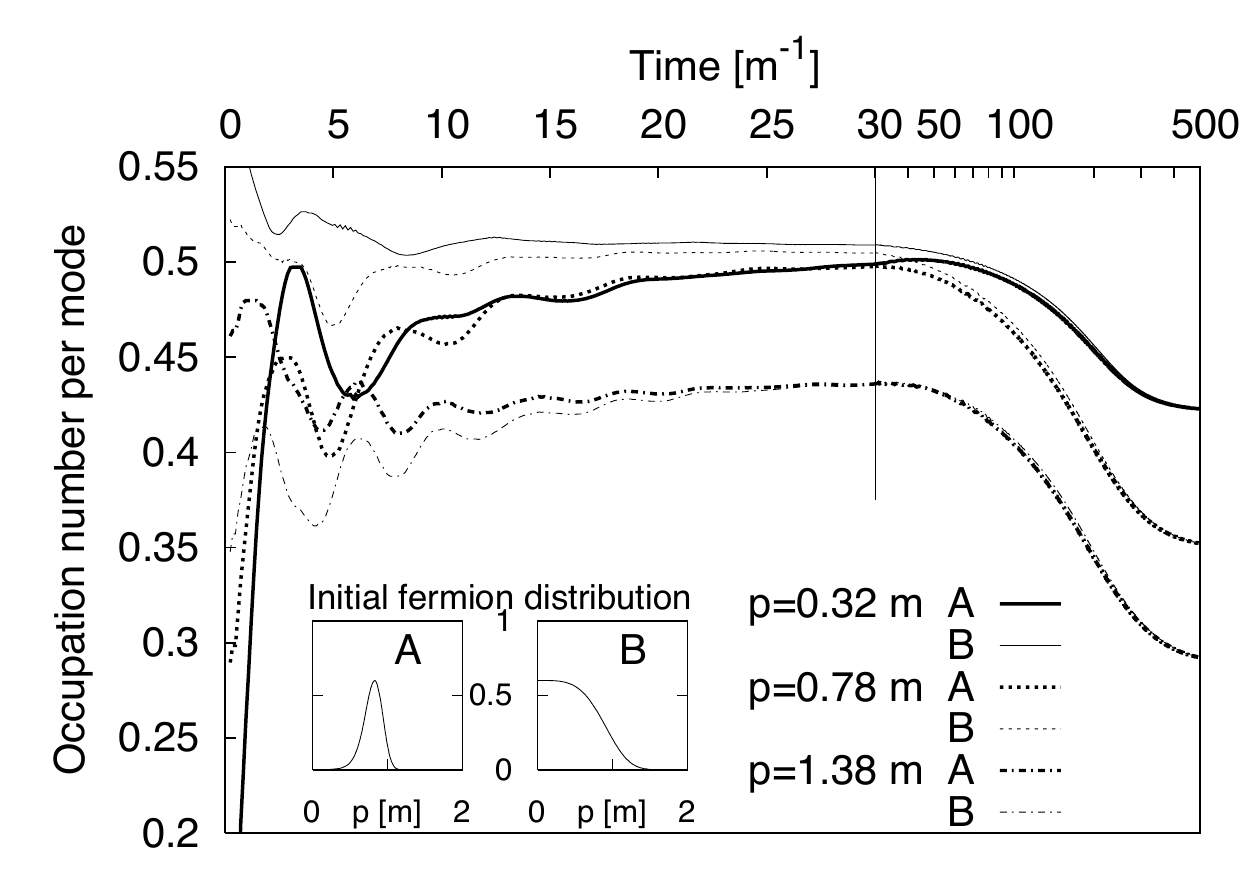}
\hspace{0.03\textwidth} 
\includegraphics[width=.43\textwidth, height=0.315\textwidth]{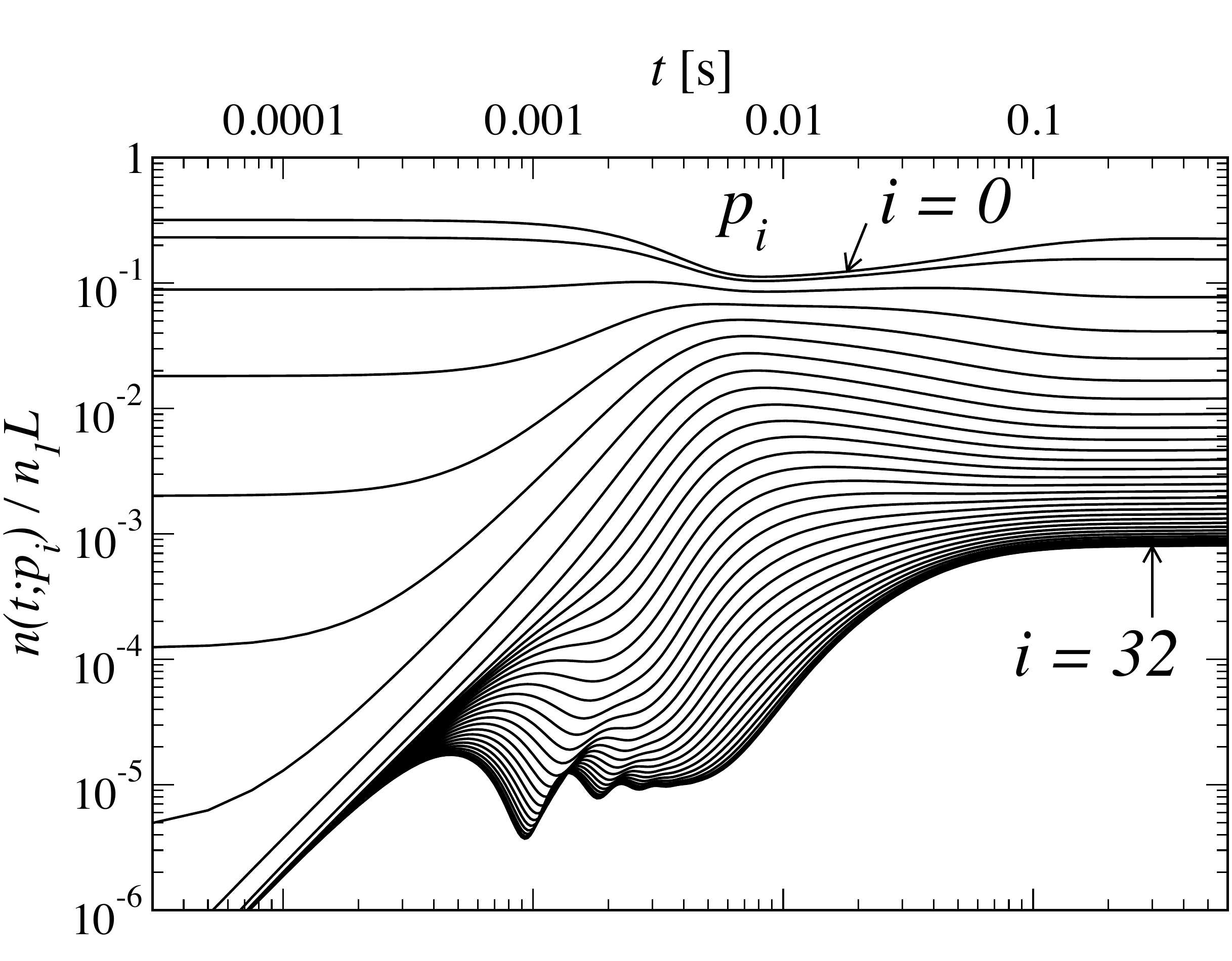}
\vspace{-0.02\textwidth}
\flushleft{\small\quad(b)\hspace{0.46\textwidth}(d)}\\[-3.5ex]
\centering
\hspace*{0.015\textwidth}
\includegraphics[width=.438\textwidth]{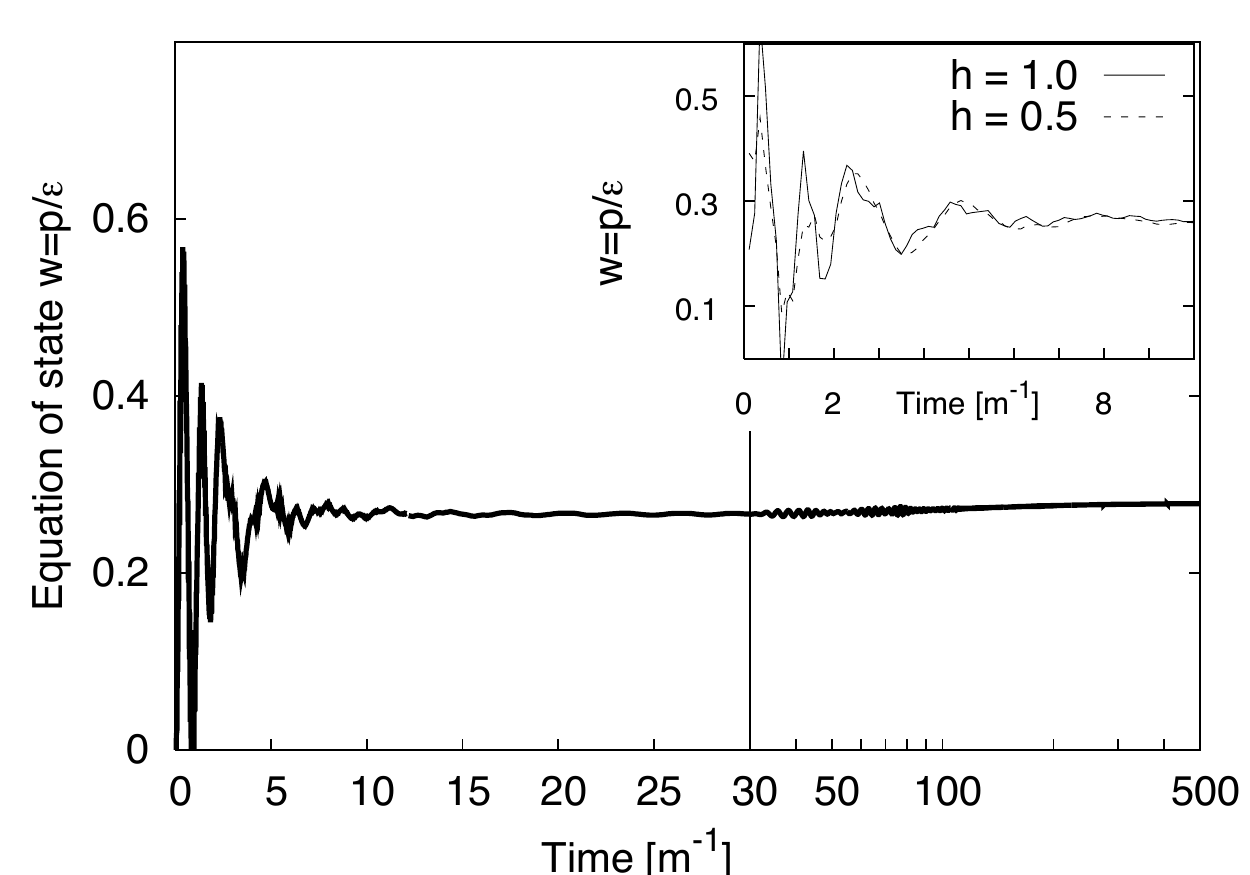}
\hspace{0.04\textwidth} 
\includegraphics[width=.435\textwidth]{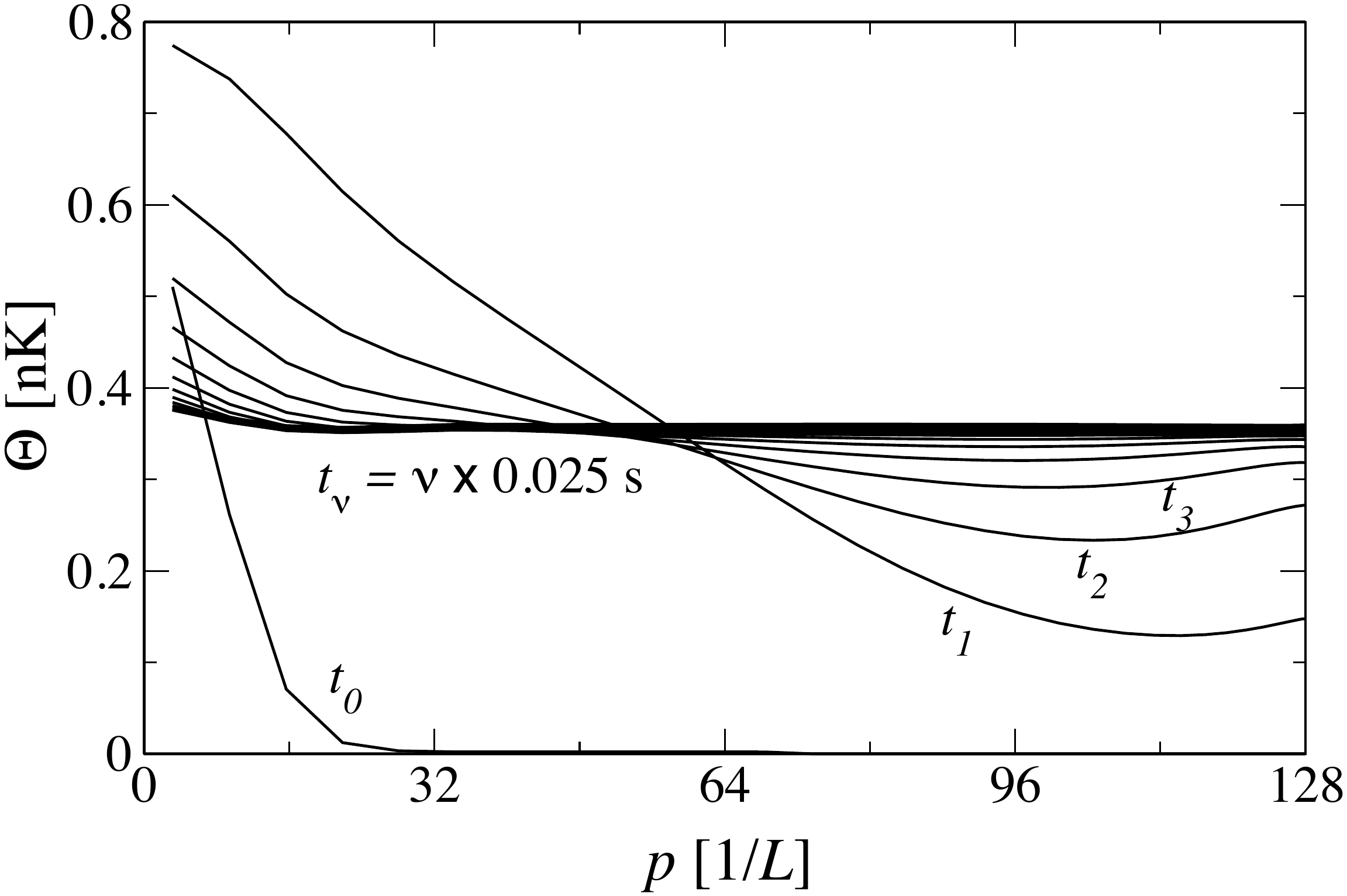}\\
\vspace*{-0.28\textwidth}\hspace*{0.7\textwidth}
\includegraphics[width=.175\textwidth]{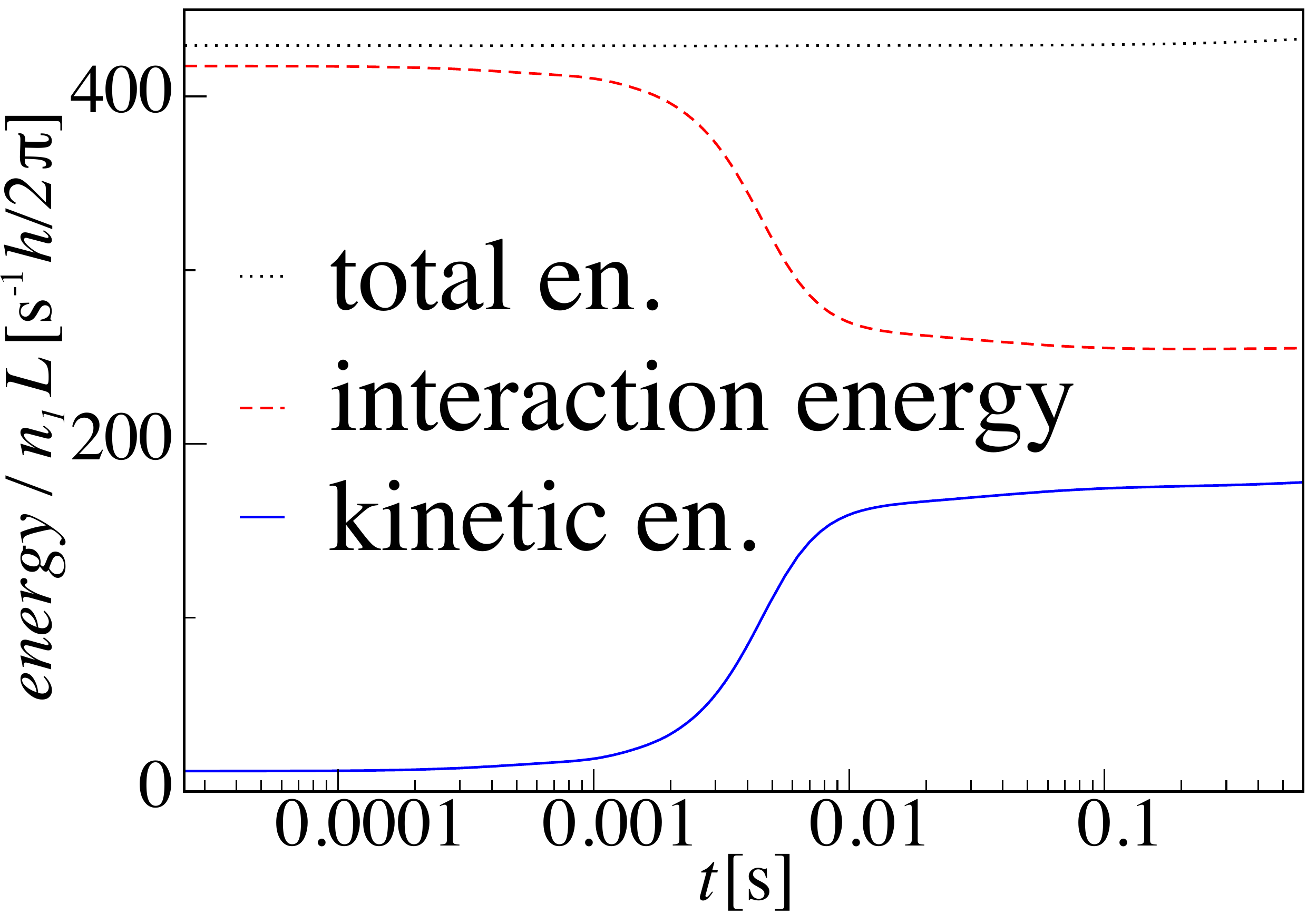}\\
\vspace*{0.17\textwidth}
\caption{Prethermalization. 
(a) Time evolution of Fermion occupation number $n^{(f)}(t;p)$ in a low-energy quark-meson model, for three different momentum modes as a function of time. 
The evolution is shown for two different initial conditions with the \emph{same} energy density. 
(b) The ratio of pressure $p$ over energy density $w$ as a function of time. 
The inset shows the early stages for two different couplings and demonstrates that the prethermalization
time is independent of the interaction details.
Adapted from \cite{Berges:2004ce}. 
Note the logarithmic scale for times $t \ge 30\, m^{-1}$. 
(c) Time evolving momentum ($p$) spectrum of a uniform dilute 1D Bose gas after an interaction quench heating the gas, shown for the lowest $32$ modes.
A fast short-time kinetic prethermalization period is followed by a long quasi-stationary drift to the final equilibrium
distribution. 
(d)
Momentum and time dependent temperature variable $\Theta(t;p)$
obtained by fitting the distribution (c) to $n(t;p)=[\exp\{(\hbar\omega(p)-\mu)/k_B\Theta(t;p)\}-1]^{-1}$, at different times $t_{\nu}$, with chemical potential $\mu$ ensuring $\omega(p)$ to be gapless.
As long as $\Theta(t;p)$ strongly depends on $p$, a generalized Gibbs ensemble applies.
Only at very large times, $\Theta$ represents the temperature of the sample.
Note that the chosen approximation introduced an integrability breaking effect.
The inset shows the energy contributions becoming stationary after prethermalization.
Adapted from \cite{Berges:2007ym}.
}
\label{fig:prethermalization}
\end{figure}

\subsection{Prethermalization}
\label{sec:prethermalization}

The statistical maximum-entropy approach does not answer the fundamental question, how the respective many-body states are actually reached given a certain microscopic model.
As discussed in the previous section, near-integrable models, which show a wide separation of relaxation time scales, are now understood to approach a prethermalized quasi-stationary state at relatively short times, long before final thermalization.
`Prethermalization' has become a collective term used for a number of at first sight apparently different relaxation phenomena.
All these phenomena have in common that they appear in the transient behavior of quantum systems relaxing from a far-from-equilibrium initial state.
The emerging intermediate state is determined by a number of conservation laws, which are relevant for the observable used and within the resolution limit of the experiment, and are usually more than total energy and particle number.

The term `prethermalization' was originally coined by Berges, Bors{\'a}nyi, and Wetterich  \cite{Aarts2000a,Berges:2004ce}, see also  \cite{Bettencourt:1997nf,Bonini1999a,Berges:2000ur} for other precursor work.
Besides the fundamental point of view these studies originated in the context of early-universe dynamics, relativistic heavy-ion collision experiments as well as the then upcoming quantum gas experiments. 

In cosmology, it is not fully resolved how particles are formed during the reheating after inflation.
In heavy-ion-collision physics, there remain discrepancies in the understanding of how a quark-gluon plasma forms out of the colliding nuclei and how the experimentally observed short equilibration times of the dense matter produced in the collision can be explained.
In view of this problem, in \cite{Berges:2004ce}, a very rapid establishment of an equation of state (pressure over energy density) was proposed to occur long before thermal, chemical equilibration sets in.
This was found to be due to bulk quantities such as the mean kinetic and potential energy equilibrating to their final thermal values much faster than, e.g., the single-particle distribution functions assume the late-time Bose-Einstein or Fermi-Dirac form.
Dephasing of initially coherently superimposed energy eigenmodes was identified as the reason of this `prethermalization'.
Three relevant time scales were identified to characterise the generic relaxation of a quantum system from a far-from-equilibrium  initial state:

(I) \emph{Kinetic prethermalization} sets in very rapidly as a consequence of the loss of phase information, i.e., of coherence between modes with different eigenfrequencies, see \Figs{prethermalization}a, c.
This dephasing is independent of the non-linear interactions between quasiparticles.
Defining a kinetic temperature in terms of the total mean kinetic energy, it is found that this temperature can take on its near-thermal value already after kinetic prethermalization, see inset of \Fig{prethermalization}d.
As the total energy is conserved, this is also true for the potential energy stored in the interactions between particles (\Figs{prethermalization}b, d).

(II) \emph{Loss of details of the initial conditions} occurs on a somewhat larger time scale, which, however, can still be much smaller than the final equilibration time.
During this period, different initial conditions lead to transient states which show similar correlations, provided conserved quantities such as energy and particle number are the same.
This applies, e.g., to mode occupation numbers of quantum gases, cf.~\Fig{prethermalization}a.
These need to be redistributed, which takes longer than dephasing, but not yet the long time needed to assume their final thermal values.
\newpage

The loss of memory generally depends on the interaction strength in a non-universal manner.
Despite the apparent relaxation, the system can be far from having reached the final equilibrium state.
If number conservation is absent and the dispersion $\omega({p})$ is known, one may define momentum-dependent mode `temperatures' $\Theta(t;{p})$.
This is done by fitting the transient correlations such as the time-dependent single-particle momentum distribution to their expected final canonical form.
For example, $n(t;p)=\{\exp[(\hbar\omega({p})-\mu)/k_{B}\Theta(t;{p})]-1\}^{-1}$ for an ideal Bose system, see \cite{Aarts2000a,Berges:2004ce} and \Fig{prethermalization}d, also for parameter definitions.
Then, a momentum dependence of $\Theta(t;{p})$ signals that final thermalization has not been reached yet.
Such a non-thermal distribution is equivalent to the system being described by a GGE.

(III) \emph{Thermalization or long-time equilibration} leading eventually to detailed balance \cite{Huang1987a,Agarwal1973a,Sieberer2015a.PhysRevB.92.134307}. 
This requires a suitable collisional redistribution within the particle spectra and can take much longer than the first two steps.
It can also be partially or completely inhibited in integrable quantum systems due to the large number of conserved quantities, and only allowed through weak integrability breaking effects.
If the system consists of more than one species of particles, \emph{chemical equilibration} can define another time scale in the overall process \cite{Berges:2004ce,Berges:2004yj}.
In the example shown in \Fig{prethermalization}c, stages II and III coincide.

As compared to the near-integrable systems discussed in \Sect{integrable}, prethermalization in the above  high-energy-physics examples seemingly does not rely on integrability of the underlying model. 
However, the same principles apply there.
The near-integrable dynamics is given by the early-time mean-field dephasing evolution of the interacting quantum fields, cf. also \cite{Moeckel2008a,Moeckel2009aAnPhy.324.2146M,Moeckel2010a.NJP.12.055016,Rosch2008a.PhysRevLett.101.265301,Kollar2011a,Marino2012a.PhysRevB.86.060408,VandenWorm2013a,Marcuzzi2013a.PhysRevLett.111.197203,Essler2014a.PhysRevB.89.165104,Nessi2014a.PhysRevLett.113.210402,Fagotti2014b,Brandino2015a.PhysRevX.5.041043,Bertini2015b,Babadi2015a.PhysRevX.5.041005,Bertini2015a.PhysRevLett.115.180601,Buchhold2015a.arXiv151003447B}. 
The fast time scale of dephasing is set by the slowest in the spectrum, i.e., by the inverse rest mass.
As long as dephasing dominates the weak non-linear interactions between quasiparticles, or, e.g., the thermalizing processes have not yet reached resonant amplification \cite{Berges:2004ce}, the dynamics is nearly Gaussian and thus nearly integrable.
This implies that the occupation number of any relevant field mode which is set to its value by the initial quench is an approximately conserved quantity.
Note that in more than one spatial dimension, \emph{isotropization} may play another important role within the early evolution starting from anisotropic initial conditions \cite{Berges:2005ai}.

Studies of the non-linear Schr\"odinger model of one-dimensional dilute Bose and Fermi gases by means of non-perturbative functional-integral techniques have shown a similar separation of time scales \cite{Gasenzer:2005ze,Berges:2007ym,Kronenwett:2010ic}.
They clearly showed kinetic pretherma\-lization as an early-time dephasing period.
After this dephasing, the total kinetic and potential energies were found to be of approximately equal size and close to their final values.
This behavior is reminiscent of the virial theorem for harmonic-oscillator type potentials and thus provides an additional signature that the system during the prethermalization stage essentially represents a series of decoupled harmonic oscillators.

First experimental studies of prethermalization were performed by Gring et al.~\cite{Gring2011a}, see also \cite{AduSmith2013a,Geiger2014a,Langen2013a.EPJST.217,Kuhnert2013a,Langen2015b.Science348.207,Langen2015a.annurev-conmatphys-031214-014548}, splitting a one-dimensional condensate into two copies and measuring differences in the subsequent dynamics of the two halves.
Details of these experiments are discussed in \Sect{experiments-near-integrable}.
In Ref.~\cite{Kitagawa2010a,Kitagawa2011a}, the dephasing kinetics of the system realized in the Vienna experiment  \cite{Gring2011a} were analysed by use of the Tomonaga-Luttinger description of the phase dynamics, demonstrating that the dephasing in this case leads to a thermal-like  state.
In \cite{Langen2013a.EPJST.217}, the prethermalization evolution of noise and correlation was modelled as an Ornstein-Uhlenbeck and thus Gaussian stochastic process.

As reviewed already in the previous section, this work sparked a lot of new theoretical activity.
This gave rise, in particular, to a number of more detailed studies on the relaxation of integrable systems 
\cite{Polkovnikov2011a,
Eisert2015a,
Gogolin2015a,
Girardeau1969a,Girardeau1970a,
Rigol2006a,
Cazalilla2006a,
Rigol2007a,
Calabrese2007a,
Barthel2008a.PhysRevLett.100.100601,
Eckstein2008a,
Gangardt2008a,
Iucci2009a,
Fioretto2010a,
Mossel2010a,
Gogolin2011a,
Pozsgay2011a,
Calabrese2011a.PhysRevLett.106.227203,
Calabrese2012a,
Calabrese2012b,
Cazalilla2012a.PhysRevE.85.011133,
Mossel2012a,
Caux2012a.PhysRevLett.109.175301,
Essler2012a.PhysRevLett.109.247206,
Gramsch2012a.PhysRevA.86.053615,
Caux2013a.PhysRevLett.110.257203,
Fagotti2013a.PhysRevB.87.245107,
Collura2013a.PhysRevLett.110.245301,
Mussardo2013a.PhysRevLett.111.100401,
Pozsgay2013a,
Fagotti2013b.1742-5468-2013-07-P07012,
Bertini2014a.JSMTE.P10035,
Fagotti2014a.PhysRevB.89.125101,
Cardy2014a.PhysRevLett.112.220401,
Mierzejewski2014a.PhysRevLett.113.020602,
Wouters2014a.PhysRevLett.113.117202,
Pozsgay2014a.PhysRevLett.113.117203,
DeNardis2014a.PhysRevA.89.033601,
Kormos2014a.PhysRevA.89.013609,
Sotiriadis2014a.1742-5468-2014-7-P07024,
Goldstein2014a.PhysRevA.90.043625,
Zill2015a.PhysRevA.91.023611,
Ilievski2015a.PhysRevLett.115.157201,
Essler2015a.PhysRevA.91.051602,
Cardy:2015xaa,
PerarnauLlobet2015a.arXiv151203823P,
Kastner2015a,
Zill:2016boi,
Ilievski:2016fdy,
Calabrese:2016xau,
Cazalilla2016a.160304252C,
Caux2016a.160304689C,
Essler2016a,
Vidmar2016a.160403990V} and in particular on the effect of weak to strong integrability breaking effects \cite{Moeckel2008a,Moeckel2009aAnPhy.324.2146M,Moeckel2010a.NJP.12.055016,Rosch2008a.PhysRevLett.101.265301,Kollar2011a,Marino2012a.PhysRevB.86.060408,VandenWorm2013a,Marcuzzi2013a.PhysRevLett.111.197203,Essler2014a.PhysRevB.89.165104,Nessi2014a.PhysRevLett.113.210402,Fagotti2014b,Brandino2015a.PhysRevX.5.041043,Bertini2015b,Babadi2015a.PhysRevX.5.041005,Bertini2015a.PhysRevLett.115.180601}.
The picture emerging from these studies is that a near-integrable system first relaxes to a `prethermalization plateau', i.e., to a metastable state described by a GGE defined by the initial state where it resides the longer the closer it is to integrability.

We note that during prethermalization, correlations are established locally, spreading through the system in a light-cone-like fashion. 
This was put forward for conformal field theories~\cite{Calabrese2006a.PhysRevLett.96.136801,Calabrese2007a}, see also the article by Calabrese and Cardy in this volume \cite{Calabrese:2016xau}.
It was shown that two-point and higher-order correlation functions of local observables, whose arguments lie in a finite region of size $l$, become stationary at a time on the order of $t\sim l/2c$ when the region fully falls into the light cone with apex centered at $t=0$.
This `horizon effect' reflects the fundamental Lieb-Robinson bound \cite{Lieb1972a} prescribing that information can spread only at a finite group velocity $c$.
The effect was studied numerically \cite{Cramer2008a,Eisert2010a} and in experiment \cite{Cheneau2012a,Langen2013b}.

In a different context, prethermalization has been potentially realized in dynamics where magnetisation domains and defects are being created following a quench in a low-dimensional spinor Bose-Einstein condensate  \cite{Sadler2006a,Vengalattore2008a,Vengalattore2010a.PhysRevA.81.053612,Guzman2011a,Kronjager2010.PhysRevLett.105.090402}.
This type of prethermalization \cite{Barnett2011a} will be further discussed in the context of non-thermal fixed points and universal dynamics in Sects.~\ref{sec:universal} and \ref{sec:QPT-Quench-Expts}.

\subsection{Non-thermal fixed points and universal dynamics}
\label{sec:universal}

The observation and description of prethermalization and GGEs leads to the question to what extent these phenomena are universal.
It can be argued that they  can be described within the more general framework of non-thermal critical states arising within a renormalization-group approach to time evolution, see \Fig{relaxation} for an illustration.

These ideas draw from the concept of universal critical scaling of correlation functions in equilibrium, which has been very successful in classifying and characterising matter near continuous phase transitions \cite{Hohenberg1977a,Goldenfeld1992a,ZinnJustin2004a,Ma2000a.ModernTheoryCritPhen}. 
Within the renormalization-group approach one basically looks at a physical system through a microscope and compares the pictures seen at different resolutions with each other, i.e., at different magnifications of the lens.
When looking for critical scaling, one takes the correlations in, e.g., the spatial patterns seen at a particular resolution and continuously changes the magnifica\-tion of the lens, watching how the correlations change.
Near the phase transition, correlations  `look' the same no matter which resolution they are observed with.

The reason for this observation is that the correlations are in fact self-similar under changes of the microscope resolution.
Shifting the resolution, set by a scale parameter $s$, to different spatial scales causes the correlation function $C(x;s)$ to be rescaled according to $C(x;s)=s^{\zeta}f(x/s)$.
Hence, whatever the resolution $s$ is, the correlations can be written solely in terms of the universal exponent $\zeta$ and the universal scaling function $f$.

If the above 'renormalization-group flow' of correlations under a change of the scale parameter $s$ does not change $C$ by any means, the scaling function $f$ must have reached a pure power-law behavior $f(x)\sim x^{\zeta}$ as is seen from the above scaling relation.
This situation is called a \emph{fixed point} of the flow.

These ideas have been extended to the time evolution from some non-equilibrium initial state, where the time $t$ takes the role of the scaling parameter $s$.
Thereby, increasing time $t$ can mean both, reducing the microscope resolution (resolving increasingly larger spatial scales) or increasing it.
A \emph{non-thermal fixed point} is a fixed point as defined above, in a time-evolution scenario where $t$ takes the role of $s$.
Functional renormalisation-group approaches for describing this real-time scaling  evolution were presented in~\cite{Gasenzer:2008zz,Berges:2008sr,Mathey2014a.PhysRevA.92.023635}.

In realistic physical situations,  a fixed point is reached only approximately such that $C(x;s)=s^{\zeta}f(x/s)$ holds but $f$ is not a pure power-law, i.e., it retains information of scales such as a correlation length $\xi$.
Near a non-thermal fixed point, one has  $C(x,t)=t^{\alpha}f(t^{-\beta}x)$, with two universal exponents $\alpha$ and $\beta$.
Hence, the correlation length would change as a power of time, $\xi(t)\sim t^{\beta}$.
The time evolution taking power-law characteristics is equivalent to critical slowing down.

Berges, Rothkopf, and Schmidt proposed that in the reheating of the post-inflationary universe non-thermal fixed points could arise, which excessively delayed thermalization \cite{Berges:2008wm,Berges:2008sr}. 
They manifest themselves in the above universal scaling behavior in time \cite{Orioli:2015dxa}, see also \cite{Berges:2012ty,Nowak:2013juc} for more recent reviews.

The theory of non-thermal fixed points extends the concepts of equilibrium and diffusive near-equilibrium renormalization-group theory (see \cite{Hohenberg1977a} for a seminal review) to the real-time evolution of far-from-equilibrium systems.  
Critical scaling phenomena in space and time are strongly reminiscent of turbulence in classical fluids \cite{Zakharov1992a,Frisch1995a} as well as superfluids \cite{Vinen2006a,Tsubota2008a}. 
For example, according to the seminal theory of Kolmogorov, eddies created in a fluid break down into successively smaller eddies until they become of the size set by dissipation of kinetic energy into heat.
This \emph{energy cascade} to smaller and smaller scales builds up a non-equilibrium steady state.

The concept of non-thermal fixed points also naturally includes relaxation dynamics which exhibits coarsening and phase-ordering kinetics \cite{Bray1994a} following the creation of defects and nonlinear patterns, e.g., in a quench across a phase transition.
In the following, this is illustrated by an example of a non-thermal fixed point approached after a quench in an ultracold Bose system.
The example shows in particular that there are more general prethermalization phenomena expected beyond the realm of near-integrable quantum systems in one spatial dimension.

\begin{figure}[t]
\centering
\includegraphics[width=0.6\textwidth]{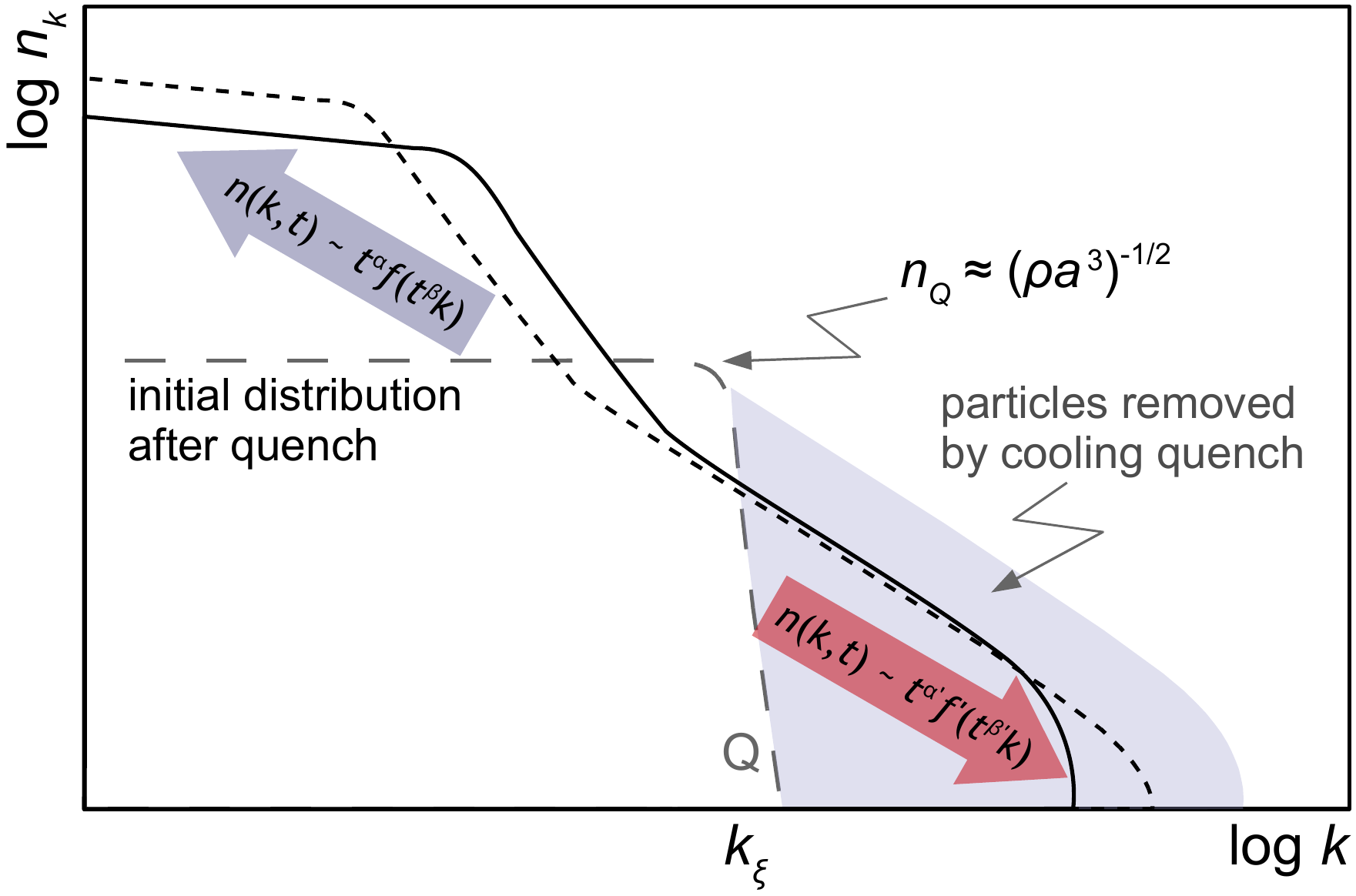}
\caption{Self-similar scaling in time and space close to a non-thermal fixed point. 
The sketch indicates the evolution of the single-particle radial number distribution $n(k,t)$ as function of momentum $k$ of a Bose gas for two different times $t$ (solid lines). 
Sketch after Ref.~\cite{Orioli:2015dxa}. 
Starting from the extreme initial distribution $n(k,t_{0})$ (dashed line) produced, e.g., by a strong cooling quench, a bidirectional redistribution of particles in momentum space (arrows) occurs. 
This builds up a quasicondensate in the infrared while refilling the thermal tail at large momenta. 
The particle transports towards zero as well as large momenta is characterized by self-similar scaling evolution in space and time, $n(k,t)=(t/t_{0})^{\alpha}n([t/t_{0}]^{\beta}k,t_{0})$, with characteristic scaling exponents $\alpha$, $\beta$, in general different for the two directions. 
The infrared transport (blue arrow) conserves particle number quasi-locally in momentum space while energy is conserved in the redistribution of short-wavelength fluctuations (red arrow).
Note the double-logarithmic scale.
In a 3D dilute Bose gas with density $\rho$, particle mass $m$, and s-wave scattering length $a$, the condition for an initial state which allows approaching the non-thermal fixed point is $(\hbar Q)^2/2m\simeq|\mu|\simeq g\rho$.
Here, $g=4\pi\hbar^{2} a/m$, i.e.~$Q\simeq k_{\xi}=(8\pi\hbar^{2} a\rho)^{1/2}$ is on the order of the inverse healing length.
Hence, if no significant zero-mode occupation $n(k=0,t_{0})$ is present, the respective occupation number at $Q$ is on the order of the inverse of the diluteness parameter, $n(Q,t_{0})\sim (\rho a^{3})^{-1/2}$. 
}
\label{fig:NTFP}
\end{figure}
Consider the dynamics following a strong cooling quench in an ultracold Bose gas, leading ultimately to the formation of a Bose condensate, see Fig.~\ref{fig:NTFP} as well as \cite{Nowak:2012gd,Berges:2012us,Orioli:2015dxa,Davis:2016hwt}.
Qualitatively this proceeds as follows.
In a closed system, a cooling quench, removing particles with the highest energies, quite generically leads to a non-equilibrated particle distribution $n(k,t)$ (dashed line in Fig.~\ref{fig:NTFP}).
This distribution, below some energy scale $\varepsilon_{0}$,  exceeds the thermal equilibrium occupation number determined by the mean energy per mode \cite{Svistunov1991a,Kagan1992a,Semikoz1995a.PhysRevLett.74.3093,Semikoz1997a}.
Energy and momentum conservation then imply a bi-directional redistribution of particles: 
while a few particles are scattered to high-momentum modes and carry away a large fraction of the excess energy associated with this over-occupation, the majority of the particles is scattered to lower momenta \cite{Nowak:2012gd,Orioli:2015dxa,Berges:2012us}.
The redistribution is indicated by the arrows in Fig.~\ref{fig:NTFP}.

Semiclassical simulations of the Gross-Pitaevskii model in three dimensions showed that this behavior is associated with the creation, dilution, coarsening and relaxation of a complex vortex tangle  \cite{Nowak:2010tm,Nowak:2011sk,Nowak:2012gd,Nowak:2013juc}, corroborating the phenomenological arguments of Ref.~\cite{Kagan1994a}.
Specifically, the shift of the infrared part of $n(k,t)$ to larger spatial scales (i.e.~smaller momenta) reflects the decay in the number of defects  \cite{Schole:2012kt,Kozik2004a.PhysRevLett.92.035301,Kozik2009a,Connaughton2005a} and thus the increase in mean inter-vortex distances.
This diluting ensemble of vortices can be considered as a type of superfluid (quantum) turbulence.

The resulting inverse transport is described by characteristic non-thermal scaling functions, separately in the low- and high-energy regions, as sketched in Fig.~\ref{fig:NTFP}.
These shift self-similarly as time proceeds (note the double-log scale).  
For the momentum-space occupation number the scaling relations in time and space translate to $n(k,t)=(t/t_{0})^{\alpha}n([t/t_{0}]^{\beta}k,t_{0})$, with universal exponents $\alpha$ and $\beta$.
For a dilute Bose gas in $d=3$ dimensions, possible scaling exponents have recently been numerically determined to be $\alpha=1.66(12)$, $\beta=0.55(3)$, in agreement with the analytically predicted values, $\alpha=\beta d$, $\beta=1/(2-\eta)$, when assuming $\eta\simeq0$~\cite{Orioli:2015dxa}.

The transport of particles towards the infrared sketched in Fig.~\ref{fig:NTFP} eventually leads to the formation of a Bose condensate.
In the presence of the rather stable turbulent tangle of vortices, this condensation proceeds, however, in critically slowed  manner, meaning that the condensate mode builds up with a smaller power-law exponent than would be the case without vortices present, as discussed in \cite{Schole:2012kt,Orioli:2015dxa}.

The above scenario of slowed relaxation in the presence of defects represents an example of phase-ordering kinetics  near a non-thermal fixed point.
Phase-ordering kinetics in general describes the growth of order through domain coarsening when a system is quenched from a disordered into a broken-symmetry phase \cite{Bray1994a}.
Thereby, a central role is taken by the dynamical exponent $z=1/\beta$.
Different such exponents, not contained in the classes considered usually \cite{Bray1994a} are found at later times of the vortex-dilution process \cite{Schole:2012kt}, indicating that the concept of non-thermal fixed points can lead beyond usual phase-ordering kinetics.

Non-thermal fixed points  \cite{Berges:2008wm,Berges:2008sr} were discussed in various contexts, including strong wave turbulence in low-energy Bose gases~\cite{Scheppach:2009wu}, in relativistic scalar models \cite{Berges:2010ez,Gasenzer:2011by}, as well as abelian \cite{Gasenzer:2013era} and non-abelian gauge theory \cite{Berges:2008mr}
They were furthermore related to classical Burgers turbulence \cite{Mathey2014a.PhysRevA.92.023635}.
Early proposals of non-thermal fixed points can be found in \cite{Bettencourt:1997nf,Bonini1999a}.

Besides vortices in 3D gases, other types of (quasi-) topological and non-topological but strongly non-linear excitations can play a role in the manifestation of the scaling.
For example, solitons can form ensembles of nonlinear quasiparticles in a one-dimensional system \cite{Schmidt:2012kw}.
In two spatial dimensions, Onsager-type \cite{Onsager1949a} ensembles of logarithmically interacting vortices and antivortices determine the  behavior near the non-thermal fixed point \cite{Schole:2012kt}. 
The latter type of systems has recently been explored extensively in experiment \cite{Weiler2008a,Neely2010a,Neely2013a.PhysRevLett.111.235301}
and theory \cite{Bradley2012a.PhysRevX.2.041001,
Reeves2012a.PhysRevA.86.053621,
Reeves2013a.PhysRevLett.110.104501,
Reeves2014a.PhysRevA.89.053631,
Billam2014a.PhysRevLett.112.145301,Simula2014a.PhysRevLett.113.165302,Reeves2015a.PhysRevLett.114.155302,
Billam2015a.PhysRevA.91.023615,
Groszek2015a.arXiv151106552G,Yu2015a.arXiv151205517Y,Salman2016a} investigating the role of quantum vortices in 2D dynamics and studying connections with aspects of two-dimensional classical turbulence.
Further examples of non-thermal fixed point scaling have been discussed for pseudo-spin and spinor gases \cite{Karl:2013mn,Karl:2013kua,Heupts2014a,Williamson2016a.PhysRevLett.116.025301}, holographic superfluids \cite{Ewerz:2014tua}, and gauge systems \cite{Gasenzer:2013era,Mace:2016svc}.
The latter indicate that such excitations may also be present in the case of turbulent prethermalization in a gluon plasma produced in heavy-ion collisions \cite{Berges:2008mr,Berges:2013fga,Berges:2013lsa,Berges:2014bba,Berges:2014yta}.  

The above non-linear excitations indicate that near non-thermal fixed points, it will be of interest to study the special role of the spectral properties encoded in the spectral or response function of the system.
Note that out of equilibrium, the spectral and statistical correlations are in general no longer related by a fluctuation-dissipation relation \cite{KadanoffBaym1995a}.
Hence, response functions can be studied in view of prethermalization and thermalization \cite{Branschadel:2008sk,Rossini2009a,Kronenwett:2010ic,Mitra2011a.PhysRevLett.107.150602,Foini2011a.PhysRevB.84.212404,Foini2012a.JStatMTE.P09011} and should give important insight also concerning non-thermal fixed points \cite{Berges:2008wm,Berges:2008sr,Scheppach:2009wu}, coarsening dynamics \cite{Chiocchetta2015a.PhysRevB.91.220302,Maraga2015a.PhysRevE.92.042151} after quenches near criticality \cite{Marcuzzi2014a.PhysRevB.89.134307}.

Turning back to 1D near-integrable quantum systems: 
Prethermalization in such systems, meaning the approach of a state well described by a GGE, represents a special case of a Gaussian non-thermal fixed point \cite{Bettencourt:1997nf,Bonini1999a,Aarts2000a}. 
This means that the effective interaction coupling of the prethermalized modes almost vanishes and the evolution essentially becomes that of a non-interacting system, usually describable as nearly free quasiparticle modes. 
Note that the conserved quantities are encoded in the non-universal scales of this partially universal state.
For any of these scales, the respective $\alpha$ and $\beta$ are close to zero such that there is no further time evolution of the scale.
When the system finally thermalizes, it departs from the  fixed point, and the time evolution can become non-universal and non-scaling.

Dynamics near non-thermal fixed points goes beyond the mere dephasing of the fundamental quasiparticle modes as discussed in the previous sections.
Non-thermal fixed points address a wider spectrum of prethermalization phenomena:
Broadly speaking, in the cases where non-linear excitations are involved, the degrees of freedom which become independent and decorrelate in the prethermalization stage, first have to emerge in the early non-linear time evolution,
cf.~also \cite{Sciolla2013a.PhysRevB.88.201110,Chandran2013a.PhysRevB.88.024306,Smacchia2015a.PhysRevB.91.205136,Chiocchetta2015a.PhysRevB.91.220302,Maraga2015a.PhysRevE.92.042151,Maraga2016a.160201763M} and \Sect{QPT-Quench-Expts}  in the context of quenches near an equilibrium critical point.
As a consequence, \emph{prethermalization} may be seen as the concept overarching all these phenomena.

We however also note that wave-turbulent cascades could be captured within the concept of a GGE, with conserved operators defined through universal scaling functions \cite{Gurarie1995a}.
In this case, GGEs have been defined which are applicable for driven stationary wave-turbulent states and thus do not take into account the self-similar time evolution governed by the exponents $\alpha$ and $\beta$.

Scrutinizing these concepts in experiment and deepening their mathematical foundations is an exciting task for future research.
Non-thermal fixed points so far have not been identified in experiment.
However, theoretical work suggests that they may have played an important role in a number of situations where the approach of a thermal state has seemingly been suppressed, e.g., in the spinor-gas experiments discussed in
\Sect{QPT-Quench-Expts}.

\section{Experiments: Relaxation dynamics in near-integrable systems}
\label{sec:experiments-near-integrable}
The experimental study of relaxation dynamics requires well-controlled and truly isolated quantum systems. In this context, ultracold neutral atoms provide unique opportunities because of the large set of available experimental methods to isolate, manipulate and probe these systems~\cite{Ketterle1999a,Bloch2008a.RevModPhys.80.885,Langen2015a.annurev-conmatphys-031214-014548}. 

In the following, we will highlight a series of experiments in which bosonic gases were strongly confined in two spatial dimensions, effectively restricting their motion to the single remaining dimension. 
This realizes a system that is very close to the integrable Lieb-Liniger gas~\cite{Lieb1963a,Lieb1963b,Cazalilla2011a}. 
The systems are not perfectly one-dimensional because the two tightly confining directions are technically still present and are felt by the atoms~\cite{Gerbier2004a,Kruger2010a}. 
Controlling the influence of the remaining transverse degrees of freedom and of the longitudinal trapping potential provides an ideal setting to study the relaxation in near-integrable quantum systems. 

We note that this restricts our discussion to a particular aspect of the large number of non-equilibrium phenomena that can and have recently been explored using cold atoms. 
For a broader overview of the rapid experimental progress in this field we refer the reader to more general reviews~\cite{Bloch2008a.RevModPhys.80.885,Polkovnikov2011a,Caux2011a.JSMTEP02023,Stamper-Kurn2013a.RevModPhys.85.1191,Eisert2015a,Gogolin2015a,Langen2015a.annurev-conmatphys-031214-014548}.

\subsection{Realizing a near-integrable system with a one-dimensional (1D) Bose gas}
The experimental realization of a one-dimensional (1D) Bose gas follows the standard procedure used for the production of Bose-Einstein condensates, employing laser cooling, trapping, and evaporative cooling \cite{Ketterle1999a}. 
In typical experiments, the cold quantum degenerate gas of atoms is harmonically confined in all three spatial dimensions. 
To create a situation where the gas is effectively 1D, a very strong anisotropy needs to be realized in a way that the transverse confinement dominates all other energy scales.
In the following, we assume the transverse confinement to be cylindrically symmetric and characterized by a trapping energy $\hbar \omega_\perp$. 
The size of the transverse ground state is $l_0 = \sqrt{\hbar / m \omega_\perp}$, where $m$ denotes the atomic mass. 
To realize a one-dimensional system, the temperature $T$ and the interaction-determined chemical potential $\mu$ have to fulfill the conditions $k_BT< \hbar \omega_\perp$ and $\mu < \hbar \omega_\perp$ \cite{Petrov2000b,Kruger2010a}.  
This is equivalent to stating that the lowest transverse excited states in the trap have a negligible occupation, i.e., that the trapped quantum gas is transversely in the ground state. 
It is interesting to note that, for a weakly interacting Bose gas, the essential criterion to achieve, $\mu < \hbar \omega_\perp$, is in fact independent of the transverse confinement scale $\omega_\perp$.
It is fulfilled whenever $2 a_\mathrm{s} n_\mathrm{1D} < 1$, where $a_\mathrm{s}$ is  the atomic $s$-wave scattering length and $n_\mathrm{1D}$ is the linear density.  
For example, in typical experiments with $^{87}$Rb ($a_\mathrm{s} = 5.3\,$nm), this requires  $n_\mathrm{1D} < 100$ atoms$/\mu$m.

Such strongly anisotropic geometries can be realized using either magnetic microtraps on atom chips \cite{Folman2002a,Reichel}, or optical dipole traps \cite{Grimm2000a}, or with optical lattices \cite{Bloch2005b}. 
The first two allow implementations of single 1D systems,  the latter realize many, nearly identical 1D systems in parallel. 

In a 1D system, all transverse degrees of freedom are effectively frozen out and excitations can only propagate
along the longitudinal, weakly confining direction.  
For bosons, this leads to markedly different behavior than in three-dimensional (3D) Bose-Einstein condensates. 
In a 3D BEC, only the lowest momentum mode is macroscopically occupied as the gas is cooled to lower and lower temperatures~\cite{Ketterle1999a}. 
In a 1D confinement, the scaling of the density of states $\rho(E) \propto \sqrt{E}$ leads to a large occupation of many momentum modes.
This is the origin of strong density and phase fluctuations, which prevent the creation of long-range order~\cite{Mermin1966a,Hohenberg1967a} and lead to a more complex set of possible equilibrium quantum states~\cite{Petrov2000b}. The key parameters determining the state of the system are the temperature $T$ and the interaction parameter $\gamma = m g_\mathrm{1D}/ \hbar^2 n_\mathrm{1D}$ where $g_\mathrm{1D}=2 \hbar^2 a_\mathrm{s} / m l_0$ is the 1D interaction strength. 
For typical temperatures $T$ reached in experiments and $\gamma \gg 1$ the system is a strongly interacting gas of hard-core bosons (Tonks-Girardeau regime)~\cite{Olshanii1998a,Dunjko2001a,Paredes2004a,Kinoshita2004a}. 
For $\gamma \ll 1 $ the gas is a weakly interacting quasi-condensate where density fluctuations are suppressed but the many occupied momentum modes lead to strong phase fluctuations.

\subsection{A quantum Newton's cradle}

The first result visualizing how integrability influences the relaxation in such a system was the experiment by Kinoshita et al.~\cite{Kinoshita2006a}. 
Atoms were trapped in an optical lattice providing strong confinement in two transverse directions, realizing a 2D array of 1D Bose gases. 
Changing the strength of the radial confinement allowed for a tuning of $\gamma$ all the way from weak to strong interactions~\cite{Kinoshita2004a}. 
By applying an optical phase grating~\cite{Cronin2009a}, a superposition of two longitudinal momentum states with opposite sign was imposed on the trapped gas. 
Given these initial conditions, the atoms started to oscillate in the trap, much like a Newton's cradle. 
These oscillations were directly imaged, and revealed a persistent non-thermal distribution. 

\begin{figure}[t]
\centering
\includegraphics[width=1\textwidth]{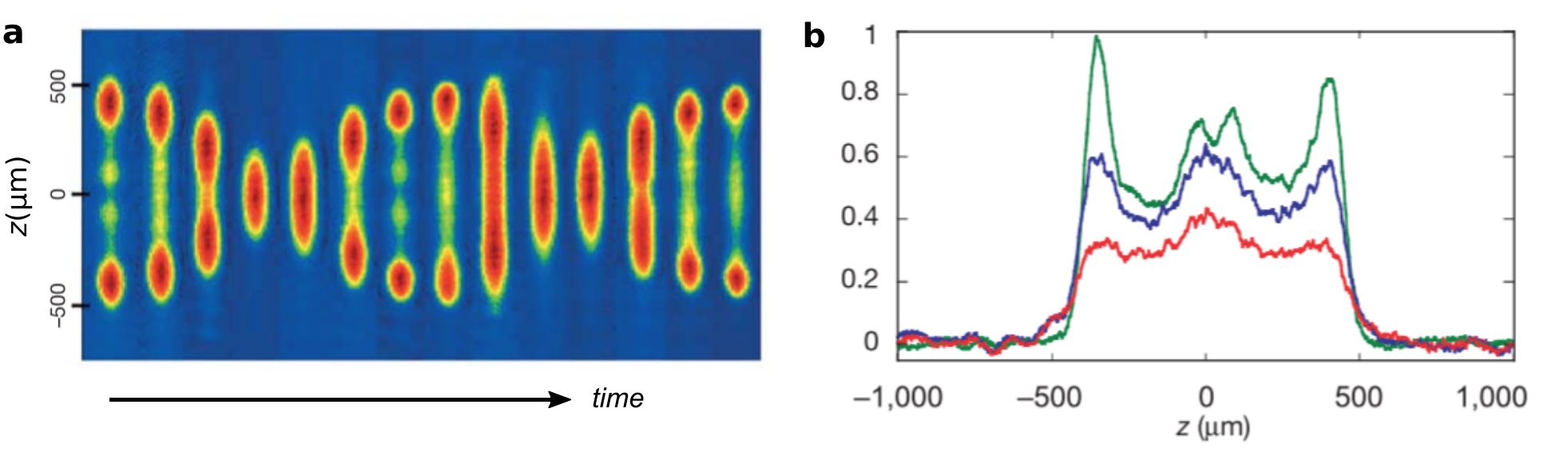} 
\caption{Quantum Newton's cradle realized with a 1D Bose gas~\cite{Kinoshita2006a}. 
(a) Long-lived oscillations in momentum space demonstrate the absence of thermalization in this near-integrable system. 
A period of $\tau=13\,$ms is shown. 
(b) Examples of expanded momentum distributions for $\gamma=4$ and three different evolution times $t=\tau=34\,$ms (green curve), $t=15\tau$ (blue), $t=30\tau$ (red), which clearly reveal a non-thermal nature. 
Figure adapted from Ref.~\cite{Kinoshita2006a}.}
\label{fig:cradle}
\end{figure}

Examples of the main observations reported in \cite{Kinoshita2006a} are shown in Fig.~\ref{fig:cradle}. 
The images provide striking evidence of the momentum distribution remaining non-thermal even after thousands of atom-atom collisions for all realized interaction strengths. 
Remarkably, when creating the same initial state in a 3D gas without the optical lattice, the system relaxed back to a thermal momentum distribution within a few collisions.
The experiment thus confirmed that integrable or near-integrable 1D systems need extremely long timescales to thermalize, while the same system thermalizes immediately when the dynamical constraints are released. 
In principle, the near-integrability should also lead to a very slow thermalization in the constraint case, in analogy with the concept of long-time equilibration introduced in Sect.~\ref{sec:prethermalization}. 
This could shed light on the fate of the KAM theorem in quantum mechanics (cf.~\Sect{integrable}). 
However, technical imperfections of the experiment that become relevant on long time scales have, so far, prevented any clear observation of this behavior~\cite{Mazets2008a.PhysRevLett.100.210403,Andreev1980a,Tan2010a.PhysRevLett.105.090404,Buchhold2015a.arXiv151003447B}.

\subsection{Relaxation and prethermalisation of a pair of 1D Bose gases}
\label{sec:SplitRelax-Expt}

The intricate microscopic dynamics that result from the interplay of relaxation, thermalization and integrability were observed in a series of experiments in Vienna~\cite{Gring2011a,Kuhnert2013a,Langen2013b,Langen2015b.Science348.207,Langen2015c}. 
In these experiments, a 1D Bose gas was created using a {\em single} magnetic microtrap on an atom chip \cite{Folman2000a,Folman2002a,Reichel}. 

Note that experiments with single systems allow for measurements that are conceptually different from the ones performed with ensembles of 1D systems in an optical lattice. 
In the latter, individual 1D gases slightly differ, e.g., in atom number, and because they are all realized and probed in parallel only ensemble averages are accessible. 
Due to the central-limit theorem one obtains Gaussian distributions. 
In contrast, only single systems allow exploring the full probability distribution functions of a quantum observable \cite{Hofferberth2008a.NatPhys.4.489,AduSmith2013a} (for theoretical descriptions see \cite{Gritsev2006a,Imambekov2006a,Kitagawa2010a,Kitagawa2011a}).
Moreover, correlation functions and cumulants which are of higher order than mean values and variances become accessible \cite{Schweigler:2015maa}.
These methods give significantly deeper insight into the underlying quantum states and can thus provide comprehensive details about the many-body dynamics and the resulting relaxed states. 

\begin{figure}[t]
\centering
\includegraphics[width=.95\textwidth]{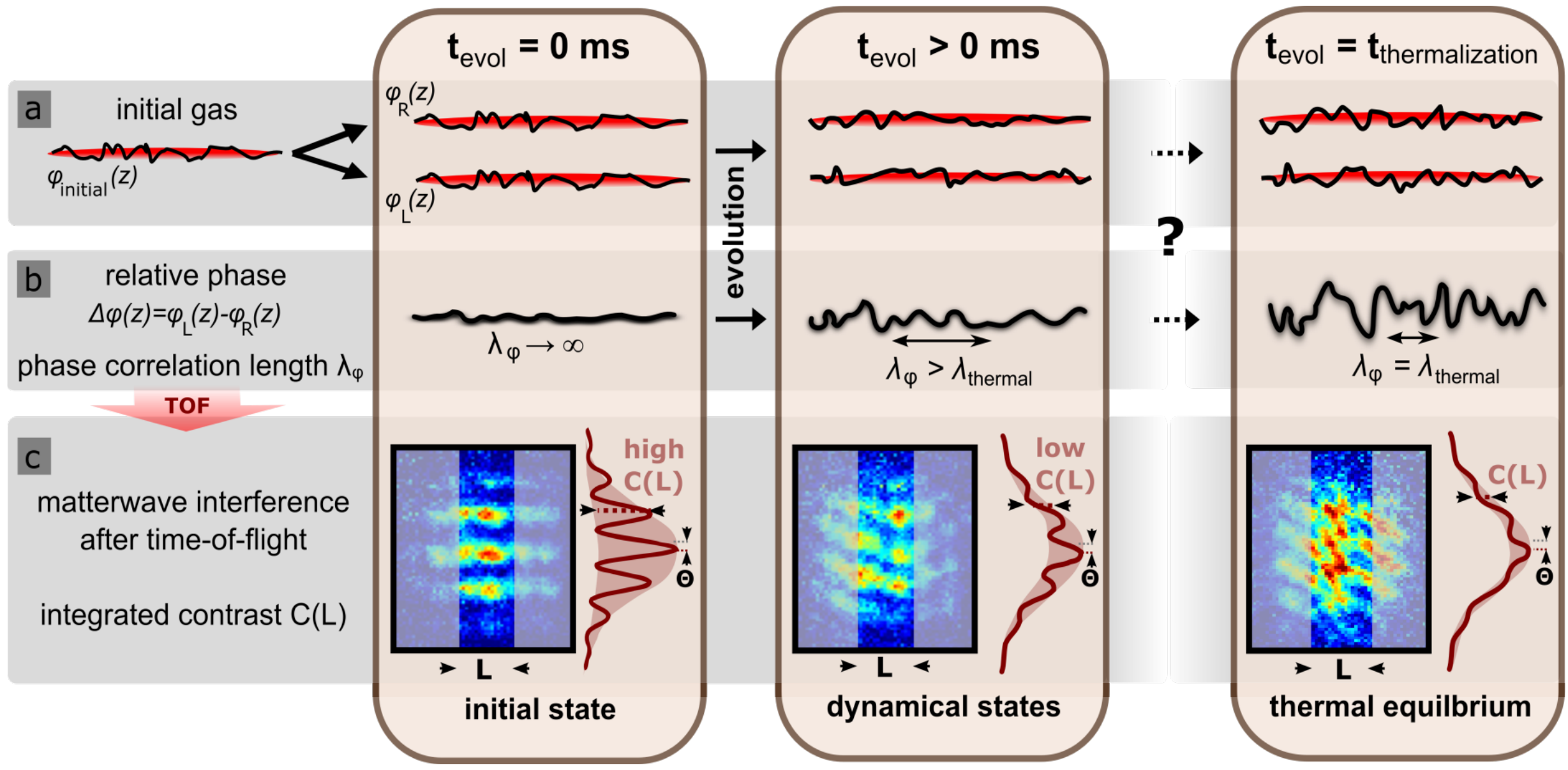} 
\caption{
Outline of the Vienna experiments described in \cite{Gring2011a,Langen2013a.EPJST.217,Geiger2014a,Langen2015b.Science348.207,Langen2015a.annurev-conmatphys-031214-014548}. 
(a,b) A phase fluctuating 1D quasi-condensate is coherently split, creating two 1D gases with almost identical phase profiles $\varphi_L(z)$ and $\varphi_R(z)$ (represented by the solid black lines). 
The gases are then allowed to evolve in the double-well potential for some time $t_\mathrm{evol}$, which leads to strong fluctuations in the local phase difference $\Delta\varphi(z)$ and a decrease of the phase correlation length $\lambda_{\varphi}$. 
The question the experiment aims at answering is whether and how this dynamical state reaches the thermal equilibrium state of two independent quasi-condensates. 
In these, the phase difference between the 1D gases fluctuates strongly, and the correlation length $\lambda_\mathrm{thermal}$ is determined by their temperature and density. 
(c) The phase difference $\Delta\varphi(z)$ between the two 1D gases is probed through time-of-flight matter-wave interference of the two gases. 
The local relative phase is directly transformed to a local phase shift of the interference pattern. 
This relative phase shift or the contrast $\mathcal{C}(L)$ of the axially integrated interference pattern can then be used as a direct measure of the strength of the relative phase fluctuations. 
Figure adapted from Ref.~\cite{Gring2011a}.}
\label{fig:Exp_general}
\end{figure}

\subsubsection{Transverse splitting of a single 1D gas }
\label{sec:Splitting}

The atom chip microtrap used in the experiments enabled a precise dynamical control over the trap. 
In this way, the initial harmonic transverse confinement could be transformed into a fully tunable double-well potential by applying strong radio-frequency (RF) dressing of the magnetic sub-states of the atoms \cite{Schumm2005a,Hofferberth2006a}. This split the gas into two almost identical halves, realizing a quench and creating a non-equilibrium state characterized by the quantum noise introduced by the splitting. 

This situation is best illustrated by analyzing a splitting process that is performed fast compared to the longitudinal dynamics so that  $t_\mathrm{split} < \xi_\mathrm{h}/c = \hbar/\mu$. 
Here, $\xi_\mathrm{h} = \hbar/mc$ is the healing length, and $c = \sqrt{\mu/m}$ is the speed of sound.  
In this limit, no correlations can build up along the axial direction, and the splitting happens independently at each point in the gas. 
The process can be intuitively pictured as a local beam splitter where each atom is independently put into the left or right half of the new system with probabilities $p_L$ and $p_R=1-p_L$, respectively. 
The corresponding probability distribution for the local number of particles $N$ on either side is therefore binomial. 
These splitting fluctuations cause a locally fluctuating interaction energy and hence set the system out of equilibrium.

\begin{figure}[t]
\centering
\includegraphics[width=.5\textwidth]{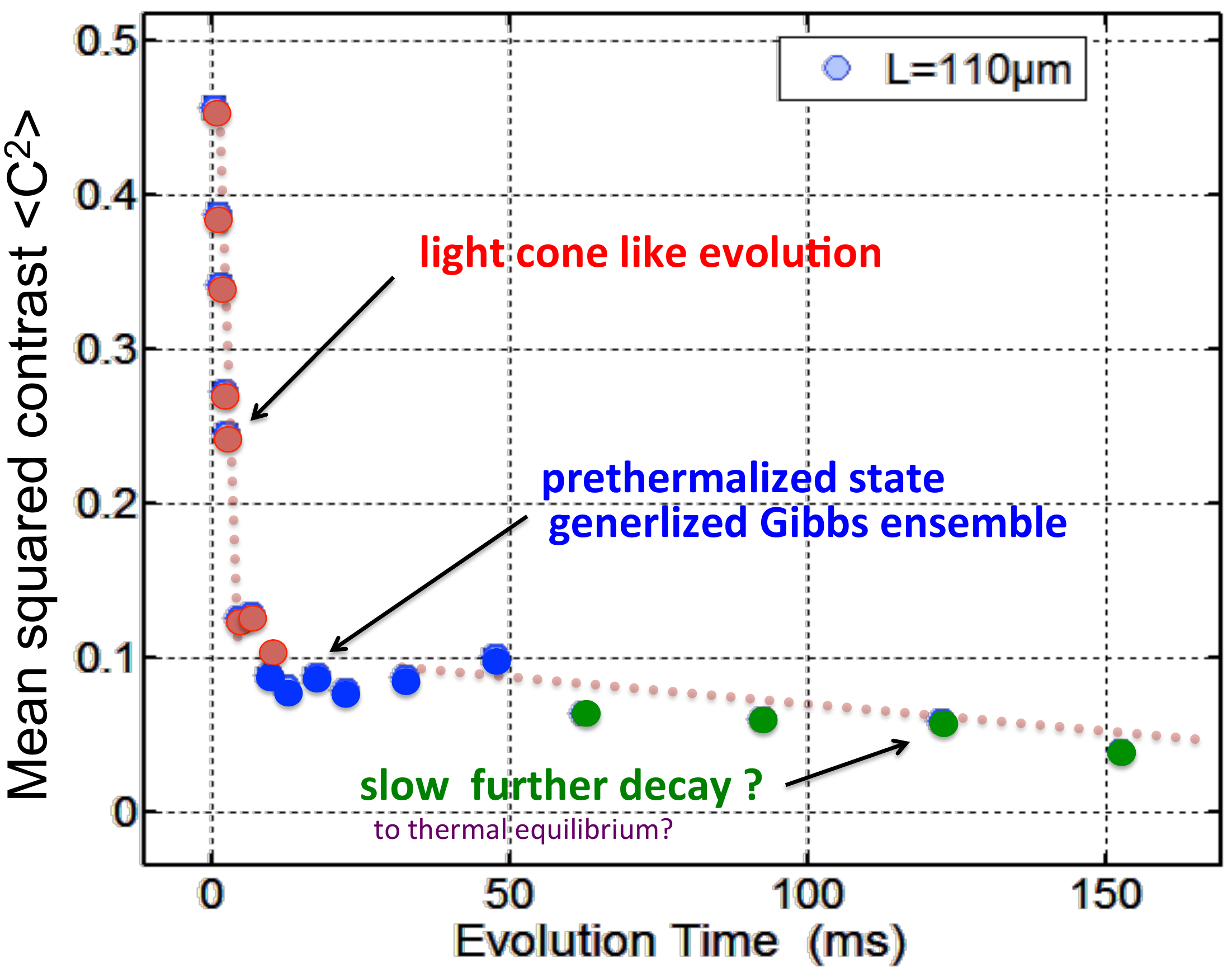} 
\caption{Relaxation behavior after splitting a single 1D quantum gas into two as revealed by the square of the interference contrast integrated over the central $110\,\mu$m of the 1D interference pattern.   
The graph shows the time evolving mean squared contrast $\langle \mathcal{C}^2\rangle$, integrated over the full length of the 1D system.
A decreasing $\langle \mathcal{C}^2\rangle$ reveals the growing fluctuations in the interference pattern as it gets more `wiggly'.  
Initially, the contrast decays quickly due to  dephasing of the approximate eigenmodes of the near-integrable Hamiltonian.  
This dephasing spreads through the system in a light-cone-like fashion~\cite{Langen2013b}, cf.~\protect\Fig{lightcone}.  
The system then relaxes towards a quasi-steady, prethermal state~\cite{Gring2011a} which is characterized by a generalised Gibbs ensemble (GGE)~\cite{Langen2015b.Science348.207}.  
The fast initial relaxation reflects the build-up of a prethermal coherence-length scale, which can be related to the fast approach of the red line on one of the blue trajectories in \Fig{relaxation}.
On longer timescales, the system shows further relaxation, and it is a key future challenge to separate and distinguish any further relaxation from processes caused by outside influences like heating due to trap instability or atom loss. 
Figure adapted from Ref.~\cite{Gring2011a}.
}
\label{fig:NonEqu_general}
\end{figure}

The outline of the experimental scheme is shown in \Fig{Exp_general}. After the splitting quench the system was let to evolve for a variable time. Subsequently, all trapping potentials were switched off, and the gases rapidly expanded transversally. This  reduced the internal interaction energy to zero on a timescale $\sim1/\omega_\perp$ such that the gas expanded mostly ballistically.  
This stopped the many-body evolution and froze the state. 
The matter waves of the two 1D gases in the double-well trap overlapped and formed a matter-wave interference pattern that could be measured by standard imaging techniques~\cite{AduSmith2011a}. 
Because of the fast reduction of the interaction energy, interactions during expansion could be neglected and the position of the fringes after time of flight approximately reflected the difference $\Delta\varphi(z) = \varphi_L(z)-\varphi_R(z)$ between the phases of the two 1D gases. The phase fluctuations can be characterized by means of the phase correlation function,
\begin{equation}
  C(z,z^\prime) 
  \sim\langle\Psi_1(z)\Psi_2^\dagger(z)\Psi_1^\dagger(z')\Psi_2(z')\rangle 
  \sim \langle e^{i \Delta\varphi (z,t)-i\Delta\varphi (z^\prime,t) }\rangle.
  \label{eq:phasecorrfunction_single}
\end{equation} 
Here, $\Psi_1$ and $\Psi_2$ denote bosonic field operators describing the two halves of the system~\cite{Langen2013b,Langen2015b.Science348.207}. 
This function provides a measure for the correlations of the phase between two different points $z$ and $z^\prime$ along the length of the system. 
It contains only the experimentally measured relative phases $\Delta\varphi(z)$ and $\Delta\varphi(z^\prime)$, and could thus be directly calculated from the data. Besides this local phase $\Delta\varphi(z)$ extracted from the relative shift of the interference pattern, a second important observable in the experiments  was the mean squared contrast $\langle \mathcal{C}^2\rangle$ of the interference patterns, integrated over a specific length scale \cite{Gritsev2006a,Imambekov2006a,Kitagawa2010a,Kitagawa2011a}.  

Repeating the experiment many times with identical initial conditions allowed studying the fluctuation dynamics of this relative phase and its relation to thermal states. 
The observed dynamics as seen through the mean squared contrast $\langle \mathcal{C}^2\rangle$ of the interference pattern is summarized in \Fig{NonEqu_general} (where the contrast $\mathcal{C}$ is not to be confused with the correlation function $C$). 
As the splitting was performed rapidly, the two halves of the system were fully coherent immediately after the splitting. 
The relative phase field was close to zero along the whole length of the system, resulting in straight fringes and thus in a large integrated mean squared contrast.  
Over time, the dynamics led to a randomization of the relative phase $\Delta\varphi(z)$, to `\emph{wiggly}' interference patterns  and a decrease of the mean square contrast $\langle \mathcal{C}^2\rangle$. 
\begin{figure}[t]
\centering
\includegraphics[width=\textwidth]{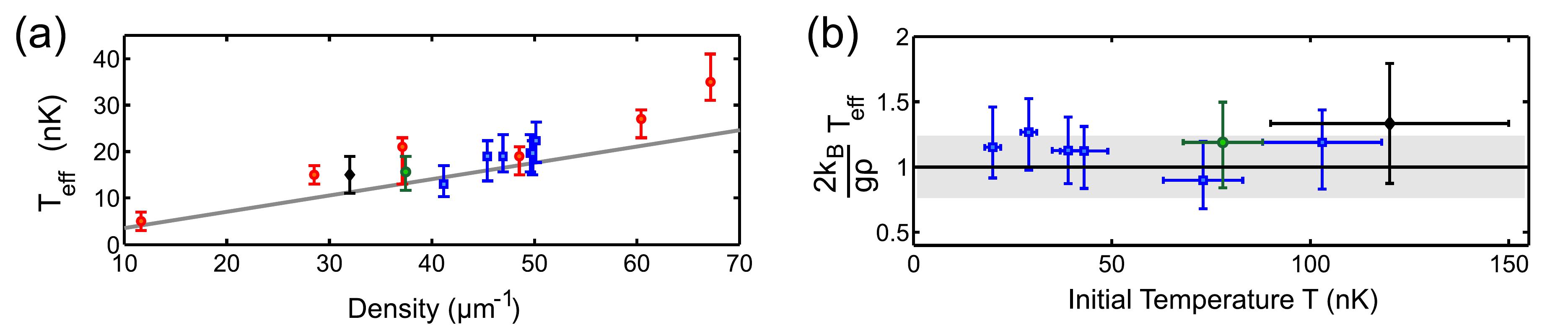} 
\caption{(a) Dependence of the effective prethermalization temperature $\Teff$ on the linear atom density $\rho$, and (b) independence of $\Teff$ of the temperature $T$ of the system before splitting, corrected for the scaling of $\Teff$ with density. 
The (black) solid line corresponds to the theoretical prediction $k_B \Teff= g n_{1D}/2$.  Figure adapted from~\cite{Gring2011a}.}
\label{fig:Exp_PreTherm}
\end{figure}

\subsubsection{Prethermalization of the relative fluctuations in a split 1D gas }
\label{sec:Pretherm-Expt}

After a sufficiently long evolution time of the pair of 1D Bose gases after the splitting, a steady state was observed. 
This state was found to be characterized by thermal full distribution functions of $\langle \mathcal{C}^2\rangle$ and exponentially decaying correlations, very much like the thermal equilibrium state for a quasi 1D Bose gas~\cite{Gring2011a,Kuhnert2013a}. However, a much lower temperature was measured than the temperature of the initial gas before the splitting.
Moreover, the longitudinal coherence length was also significantly larger than the expected thermal decay length. 
The relative degrees of freedom of the system were thus looking thermal, but with an effective temperature $T_\mathrm{eff}$ that was significantly lower than the initial temperature $T_\mathrm{in}$.

Concomitant theoretical studies~\cite{Kitagawa2010a,Kitagawa2011a} revealed  the observed extremely low temperature to be a result of the particular quench that was performed. 
Splitting the system creates new degrees of freedom the canonical coordinates of which are given by the local phase difference and the relative atom number between the two systems. 
While the individual halves of the system still fluctuate strongly with the initial temperature $T_\mathrm{in}$, only a very small amount of energy is introduced into the relative degrees of freedom via the quantum shot noise of the splitting process. 
The fast quench, in particular, leads to equipartition of this energy $\varepsilon_\mathrm{split}=g n_{1D}/2$ over the modes of the new relative degrees of freedom. 
Dephasing then establishes the thermal-like state with temperature $T_\mathrm{eff}=\varepsilon_\mathrm{split}/k_B = g n_{1D} / 2 k_B$. The theoretical model thus predicts that the effective temperature should be independent of the initial temperature, and exhibit a linear scaling with the 1D density. This behavior was indeed observed in the measurements and is shown in  \Fig{Exp_PreTherm}. It represents precisely the kinetic prethermalization \cite{Aarts2000a,Berges:2004ce} that is discussed in Sect.~\ref{sec:prethermalization}. 

Note that in the experiments, the quench introduced white-noise fluctuations into the relative degrees of freedom. This leads, within good approximation, to a GGE with only one Lagrange multiplier corresponding to the inverse of the effective temperature, $\beta_\mathrm{eff}=(k_{B}T_\mathrm{eff})^{-1}$. 
This GGE thus takes the form of a genuine Gibbs ensemble, with
$\varepsilon_\mathrm{split}\simeq\omega_{k}n_{k}(0)\simeq$ const., independent of wave number $k$, where $n_{k}(0)$ is the mode occupation number right after the quench.

\subsubsection{Light-cone-like spreading of phase decoherence }
\label{sec:LightCone-Expt}

\begin{figure}[t]
\centering
\includegraphics[width=.47\textwidth]{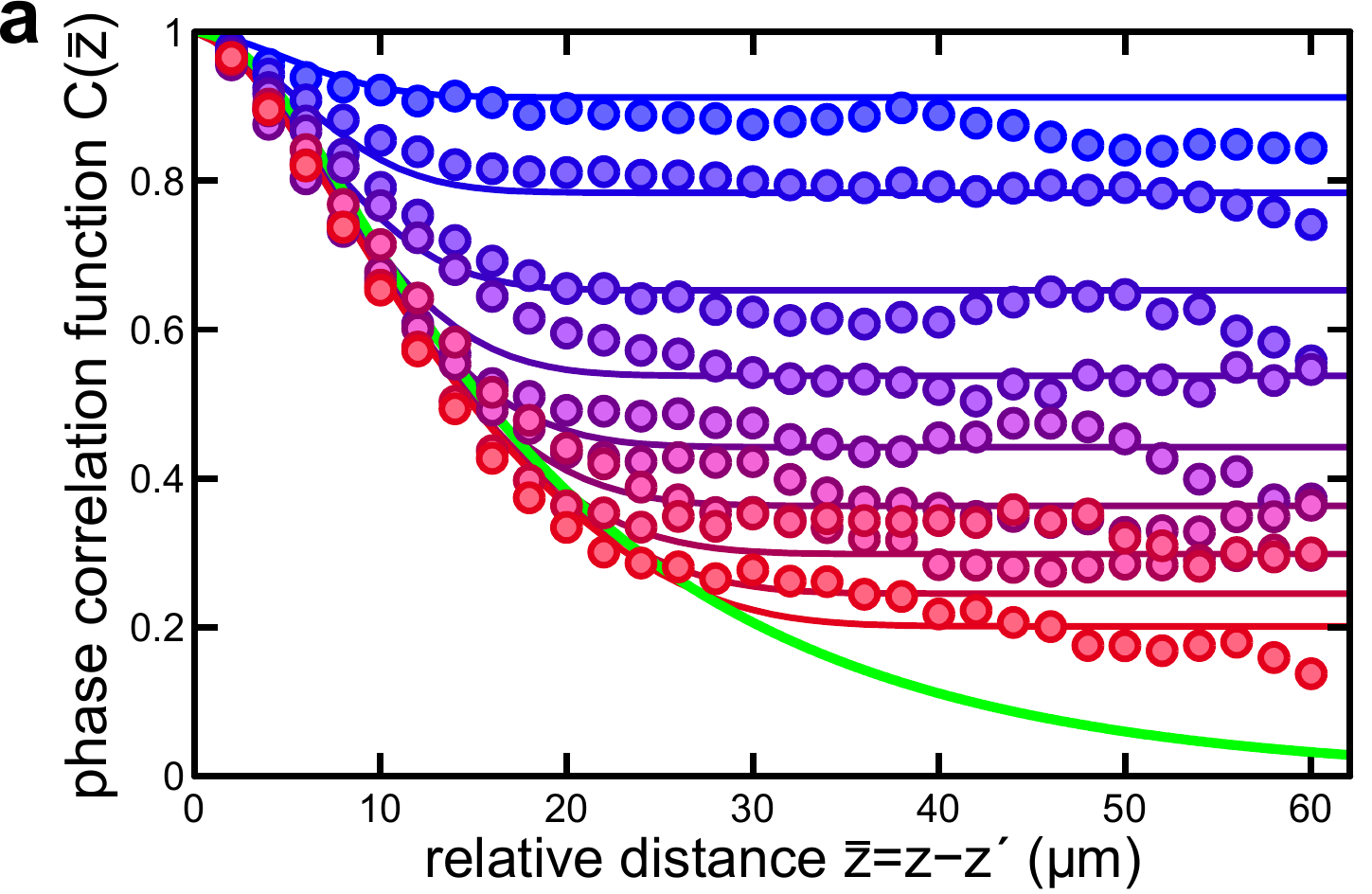} 
\includegraphics[width=.47\textwidth]{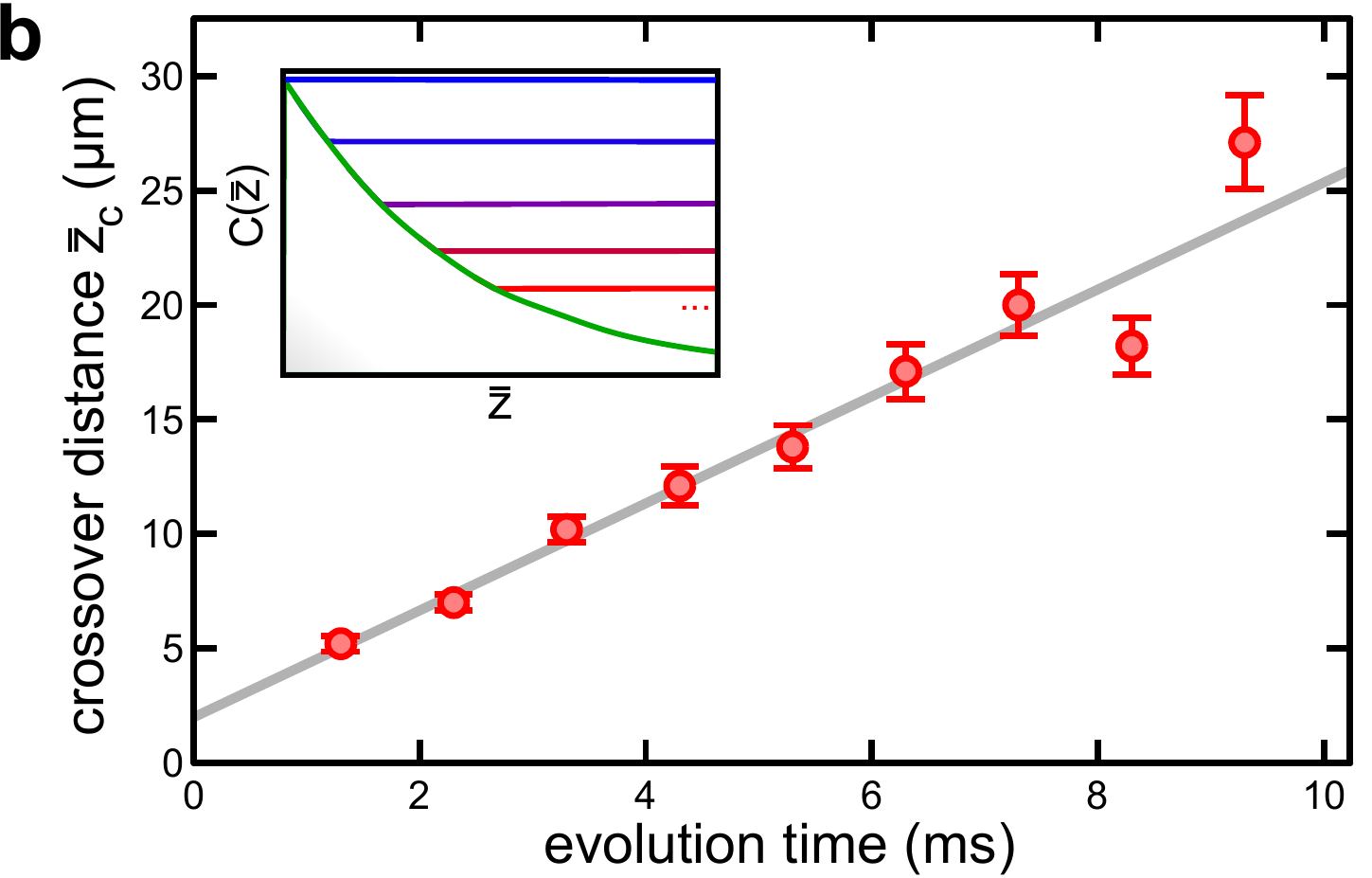} 
\caption{Temporal spreading of the prethermalized fraction. 
(a) Experimental phase correlation functions $C(\bar z,t)$ (filled circles) compared to theoretical calculations (solid lines), as a function of the relative distance $\bar z=z-z'$ between two longitudinal positions. 
The evolution time $t$ increases from top to bottom.
The final (green) line is the theoretical prediction for the relaxed, fully prethermalized state. 
At each time $t$, correlation functions follow this prediction up to a crossover distance $\bar z_c(t)$ beyond which the system remembers the initial long-range phase coherence. 
(b) Position of the crossover distance $\bar z_c$ as a function of  $t$, revealing the light-cone-like emergence of the thermal correlations of the prethermalized state. 
The slope of the solid line corresponds to twice the phonon velocity of the system. 
Figure adapted from Ref.~\cite{Langen2013b}.}
\label{fig:lightcone}
\end{figure}

An important question is how the above prethermalized quasi-steady state is established after the quench. In the experiments, the two-point phase correlation function $C(z,z^\prime)$  (see Fig.~\ref{fig:lightcone}) allowed for a detailed study of these dynamics~\cite{Langen2013b,Langen2013a.EPJST.217,Geiger2014a}. 
Directly after the splitting the correlations were close to $C(z,z^\prime)\equiv1$, which reflected the full coherence of the relative phase.  
For any given evolution time $t$ after the splitting, $C(z,z^\prime)$ revealed that the system had already established the prethermalized correlations up to a distance $z-z' = 2\, c\, t$, with the speed of sound $c$. 
At larger separations of $z$ and $z'$, the system still retained the initial long-range order imposed by the quench. 

This demonstrates how the thermal correlations of the prethermalized state were established locally and then spread through the system in a light-cone-like fashion. 
Hence, the phononic excitations of the system could be interpreted as information carriers that propagate correlations through the system. 
This reflects the basic Lieb-Robinson bound limiting the spreading of information to a finite group velocity, as originally introduced for lattice spin models \cite{Lieb1972a} and studied both numerically \cite{Cramer2008a,Eisert2010a} and experimentally \cite{Cheneau2012a}.
The limits observed here extend these ideas to continuous models, as previously put forward for quenches in conformal field theories~\cite{Calabrese2006a.PhysRevLett.96.136801,Calabrese2007a}.
\begin{figure}
\centering
\includegraphics[width=0.8\textwidth]{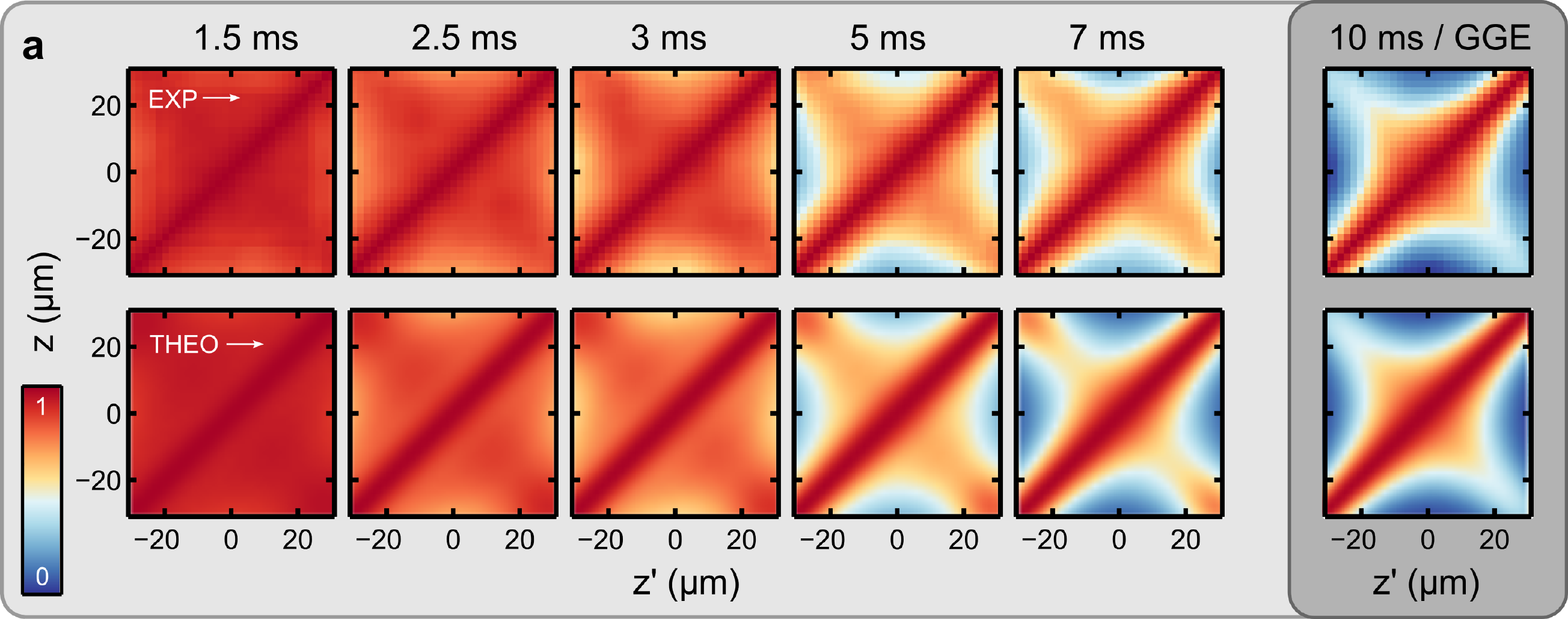} ~\\~\\
\includegraphics[width=0.8\textwidth]{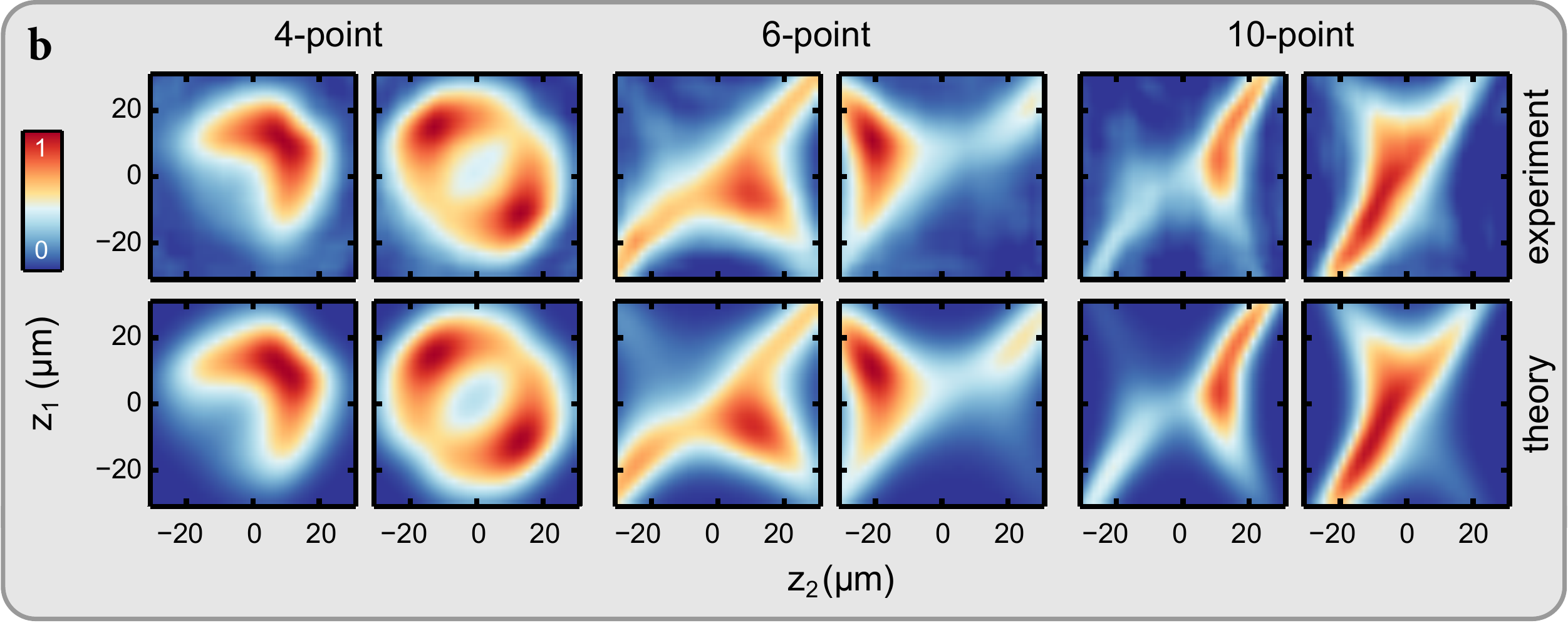} 
\caption{Observation of a generalized Gibbs ensemble (GGE).
(a) Two-point phase correlations visualizing the emergence of the GGE. 
The correlation functions show a characteristic maximum on the diagonal and a decay of correlations away from the diagonal (cf.~\Fig{lightcone}). 
Additional correlations on the anti-diagonal are the result of the enhanced occupation of even (with respect to the longitudinal trap) quasiparticle modes, in good agreement with a theoretical model assuming multiple conserved quantities. 
(b) Examples of cuts through experimental 4-, 6-, and 10-point correlation functions. 
The GGE describes well also phase correlations up to $10^\mathrm{th}$ order. 
From left to right, we plot the cuts $C(z_1,10,z_2,10)$, $C(z_1,-12,z_2,14)$, $C(z_1,10,10,z_2,-20,10)$, $C(z_1,-8,8,z_2,-24,-20)$, $C(z_1,4,10,z_2,-8,z_2,-22,-18,10,-4)$ and $C(z_1,-22,-8,z_2,-22,-26,-22,z_2,-26,$ $-24)$. 
All coordinates are given in $\mu$m and were chosen as representative cases for our high-dimensional data. 
Adapted from Ref.~\cite{Langen2015b.Science348.207}.}
\label{fig:GGE}
\end{figure}


\subsubsection{Observation of a generalized Gibbs ensemble }
\label{sec:GGE-Expt}
The experimentally observed prethermalized state can be described by a single effective temperature $T_\mathrm{eff}$. 
Although this state was observed in a system that is very well described by an integrable model, this state is practically indistinguishable from a thermal state. 
As mentioned above, the key to this surprising result lies in the particular quench protocol that was employed.
According to this protocol, all conserved quasiparticle modes of the relative degrees of freedom were prepared with the same energy $\varepsilon_{\mathrm{split}}\simeq k_B T_\mathrm{eff}$. 

\begin{figure}
\centering
\includegraphics[width=.5\textwidth]{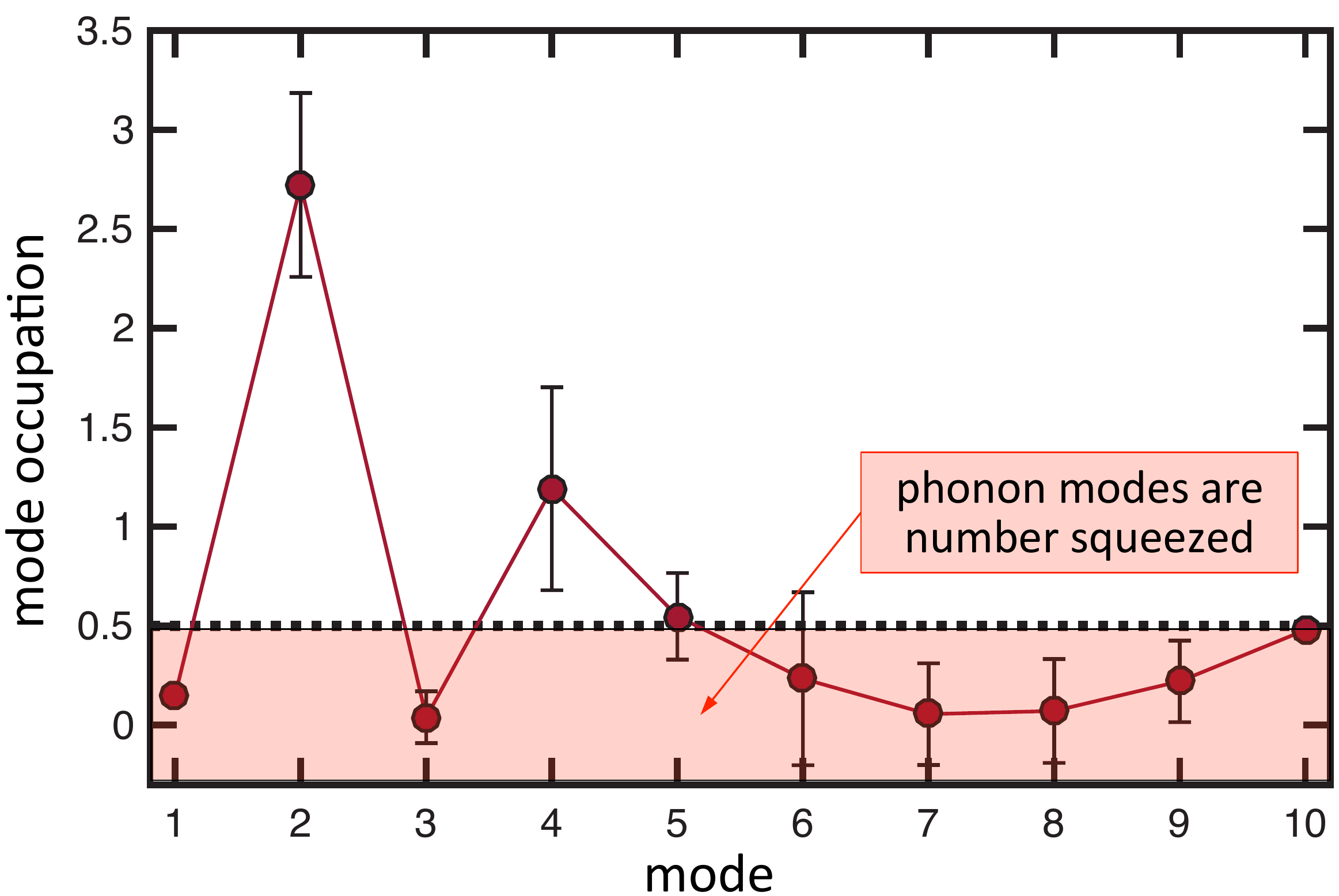} 
\caption{The generalized Gibbs ensemble (GGE) observed in \cite{Langen2015b.Science348.207}.
Shown are the mean occupation numbers $n_m$ (in units of $\varepsilon_m/\mu$) of the quasiparticle modes with index $m$ that define the GGE shown in \Fig{GGE}.  
The distribution reveals that the occupation of the lowest even (odd) modes are increased (decreased) as compared to the quantum noise occupation found in a single-temperature prethermalized state observed after an instantaneous quench (dashed line).  
Modes with occupation number below this `quantum-noise' line are number squeezed. 
Figure adapted from Ref.~\cite{Langen2015b.Science348.207}.}
\label{fig:Exp_GGEmodes}
\end{figure}

In a subsequent experiment, Langen et al.~\cite{Langen2015b.Science348.207} modified this quench protocol  by changing the speed with which the double well was established during the splitting process.
They observed a relaxed state, which showed enhanced correlations outside the light cone, strikingly visible as a cross in the non-translation-invariant phase correlations $C(z,z^\prime)$ reproduced in \Fig{GGE}a.  
A detailed analysis showed that, in fact, not less than ten Lagrange multipliers, describing the occupation of the ten lowest-energy quasiparticle modes of the relative degrees of freedom, were necessary for a precise description of the observed correlations (\Fig{Exp_GGEmodes}).  
It is interesting to note that a series of these modes, especially $n=1$ and $3$ and, to a lesser extent, $n>5$, had occupations that were significantly below the quantum-noise limit of a fast quench.  
This could be due to the fact that the employed quench created a non-trivial many-body quantum state with number-squeezed excitations. 
Studying these will be part of future research.

Theoretically, the GGE has been predicted to be applicable to describe local correlation functions of prethermalized states defined in a finite-size region, see the discussion in \Sect{gge}.
Moreover, since the GGE represents the optimum ensemble on the basis of available information, deviations of relaxed states from the GGE description are generally expected to become manifest first in higher-order correlation functions. 
Basic phase correlation functions up to $10^{\mathrm{th}}$ order have been measured \cite{Langen2015b.Science348.207}, and arbitrarily chosen two-dimensional cuts through these are shown in \Fig{GGE}b. 
Similarly to the two-point correlation functions, they were found to be in very good agreement with the theoretical model.
As demonstrated in \cite{Langen2015b.Science348.207}, the specifically varying mode occupations and thus Lagrange multipliers are necessary to correctly describe the measured correlations, in particular at high orders.
The GGE experiment thus gave a first glimpse on techniques for measuring high-order correlations, which have been developed further thereafter \cite{Schweigler:2015maa} and become useful and likely necessary for detecting complex non-local properties of dynamically evolving near-integrable quantum systems.
We point out that higher-order (connected) correlations, i.e., cumulants, are generically expected to become important in systematic studies of many-body properties including generalized Gibbs characteristics, anomalous scaling and universality.

\subsubsection{Long-time evolution and revivals }
An outstanding question concerns the further evolution after the establishment of the prethermalized state described by a GGE. 
Will the system further relax and eventually reach thermal equilibrium? 
The slow decay of the mean squared contrast $\langle \mathcal{C}^2\rangle$ over long limes seen in \Fig{NonEqu_general} hints at such a further relaxation. 
It has been predicted that higher-order scattering processes \cite{Andreev1980a,Mazets2008a.PhysRevLett.100.210403,Tan2010a.PhysRevLett.105.090404} will induce this relaxation, and numerical calculations \cite{Stimming2011a.PhysRevA.83.023618} gave further evidence. 
Experimentally, this is a very difficult problem, because the smallest heating due to, e.g., a shaking of the trap can mimic the respective processes.  
Further experimental investigations are currently under way.
   
An interesting further aspect concerns the effect of the finite size of the studied system, in particular the possibility of coherent revivals at finite times.
For quenches in conformal field theories, partial as well as full revivals were predicted to occur at integer multiples of the system size divided by twice the speed of sound, and relations to the formation of black holes existing through the AdS-CFT correspondence were pointed out \cite{Cardy2014a.PhysRevLett.112.220401,Cardy:2015xaa}.
A detailed study of the light-cone-like propagation of excitations and the build-up of a dephased, prethermalized state after the initial splitting quench gave concrete predictions that revivals should be observable in box-like potentials, but are suppressed by the harmonic confinement of the current experiments \cite{Geiger2014a}. 
Both of these problems will be at the center of future developments and experiments. 

\subsection{Implications from prethermalization for the preparation of a 1D Bose gas}
The relaxation to a prethermalized instead of a thermal state also has important implications for the preparation of the initial pre-quench gas in such experiments. 
The standard technique to prepare these gases at ultracold temperatures is evaporative cooling, which relies on the selective removal of the most energetic particles from a trapped gas and the subsequent rethermalization of the gas to a lower temperature through elastic collisions. 
For efficient cooling this cycle is being repeated continuously, increasing the phase-space density of a gas at the cost of reducing the total number of atoms. 
However, in a pure 1D setting, two-body collisions do not lead to a redistribution of energy and momentum and thus, intuitively, such systems should not be cooled through particle dissipation. 

In principle, the known conditions that break integrability in these systems could be an explanation for the observed cooling efficiency deep into the 1D regime. 
For example, for realistic confinement strengths, the 1D condition and thus the integrability of the system can be broken in collisions where there is enough energy available to access transverse excited states. 
However, these thermalizing two-body collisions are suppressed by at least a factor of $\exp(-2 \hbar \omega_\perp/k_B T)$ in a non-degenerate bosonic gas~\cite{Mazets2008a.PhysRevLett.100.210403}. 
Consequently, these collisions freeze out as soon as the gas enters the 1D regime. Other higher-order processes that can lead to thermalization are three-body collisions~\cite{Mazets2008a.PhysRevLett.100.210403,Tan2010a.PhysRevLett.105.090404} or phonon-phonon scattering~\cite{Andreev1980a,Stimming2011a.PhysRevA.83.023618}. 
However, their expected thermalization time scales are much larger than the cooling time in typical experiments.

\begin{figure}
\centering
\includegraphics[width=0.6\textwidth]{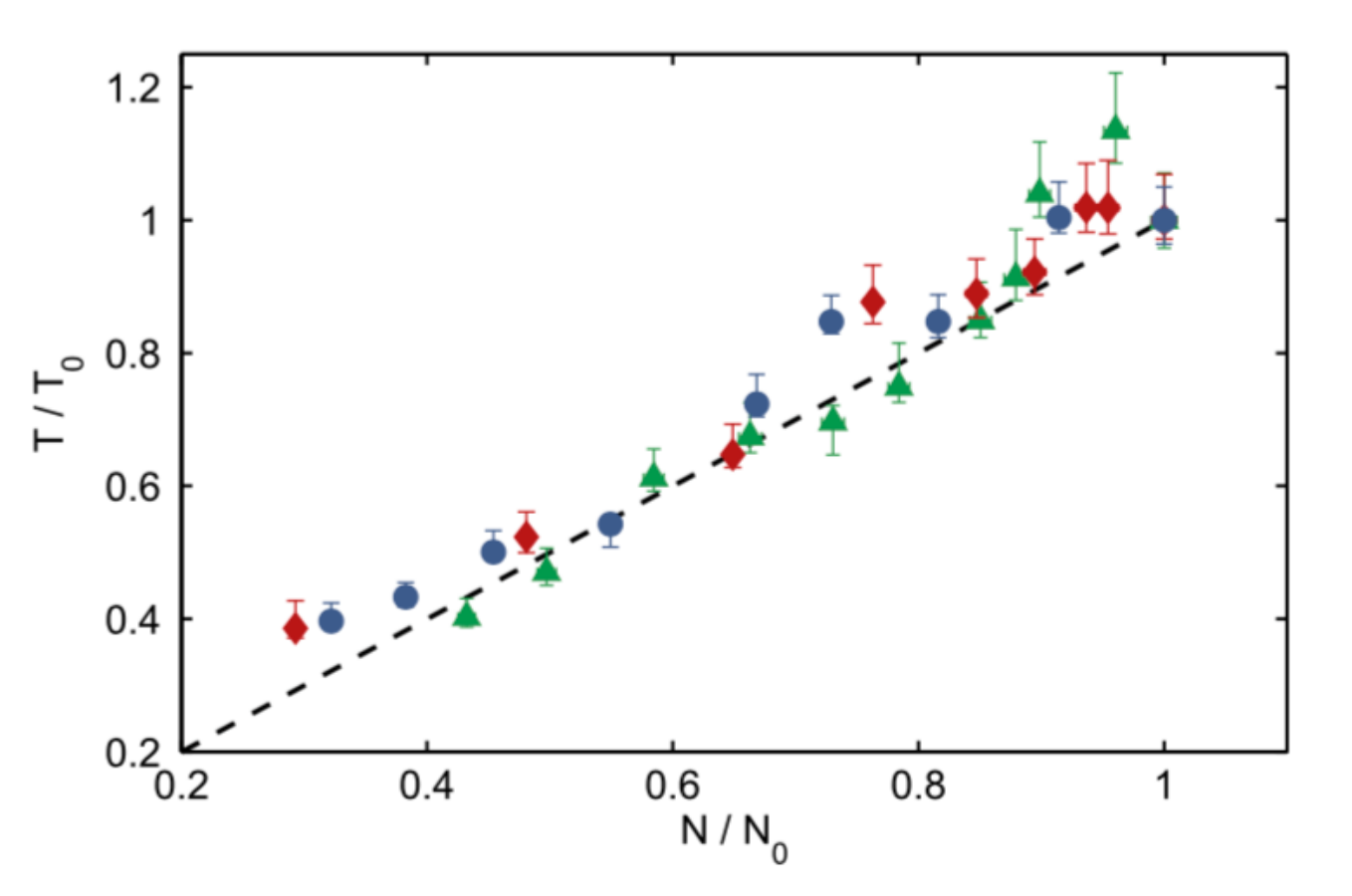} 
\caption{Using prethermalization for the preparation of a 1D Bose gas.
Shown is the universal scaling relation between temperature and atom number observed in the dissipative cooling of a near-integrable 1D Bose gas. Figure adapted from Ref.~\cite{Rauer2016a}.}
\label{fig:cooling}
\end{figure}

As in the case of prethermalization the key to understanding the cooling process of the gas is dephasing. 
In an experiment by Rauer et al.~\cite{Rauer2016a}, the outcoupling process was revealed to be nearly homogenous and independent of the mode energy. 
While this would, as discussed above, not lead to cooling, a significant decrease in temperature down to $k_B T \sim 0.1\, \hbar \omega_\perp$ was observed. 
In particular, all correlation functions remained close to their thermal form for all times despite the constant removal of atoms. 
 
The dynamics resulting from such a homogeneous particle dissipation can be intuitively understood within a Luttinger-liquid picture, describing the low-energy dynamics of the underlying Lieb-Lininger model~\cite{Grisins2016a.PhysRevA.93.033634}. 
Each of the phonon modes in this model contributes to the fluctuations in the gas through a density and a phase quadrature, in  analogy to the position and momentum quadratures of a harmonic oscillator. 
The free evolution of such a mode $k$ with energy $\hbar\omega_k$ can be visualized as a rotation of the corresponding Wigner function with the frequency $\omega_k$. 
In this picture, a sudden homogeneous outcoupling of atoms leads to a decrease in average density with the density fluctuations around this average being scaled down correspondingly. 
This instantaneous density reduction therefore extracts energy from the density quadrature of the phonon modes while leaving the phase quadrature unchanged. 
The system reacts to this reduction by dephasing, redistributing the remaining energy between the quadratures. 
This behavior is  very similar to the one observed during prethermalization. 
Both the experiment and a simple theoretical analysis established a universal scaling relation between temperature and particle number which is shown in \Fig{cooling}.

\subsection{Prethermalization after quenches to or across a critical point}
\label{sec:QPT-Quench-Expts}
The experiments discussed in the previous subsection employed an entirely 1D cooling process that suppressed excitations of the gas. 
The opposite case of a strong cooling quench \emph{into} a single 1D trap can also be realized and is expected to show similarities to quenches to or across a critical point.
In a different context, quenches across a critical point were realized in the experiments of the Stamper-Kurn  \cite{Sadler2006a,Vengalattore2008a,Vengalattore2010a.PhysRevA.81.053612,Guzman2011a}, Sengstock \cite{Kronjager2010.PhysRevLett.105.090402}, and Oberthaler \cite{Nicklas2011a,Nicklas:2015gwa} groups with spinor gases, see \cite{Stamper-Kurn2013a.RevModPhys.85.1191} for a recent review.
Quenches leading to solitary excitations, studying Kibble-Zurek-type physics \cite{Kibble1976a,Zurek1985a,Laguna1997a.PhysRevLett.78.2519,Damski2010a.PhysRevLett.104.160404,Davis:2016hwt}, were performed in experiment by Lamporesi, Ferrari, et al.~\cite{Lamporesi2013a},  Schneider, Bloch and collaborators \cite{Braun2014a.arXiv1403.7199B}, the Dalibard group \cite{Corman2014a,Chomaz2015a}, and the Hadzibabic group \cite{Navon2015a.Science.347.167N}.
All of these contributed towards studying dynamical critical scaling and universality.

Employing semiclassical Truncated-Wigner-type simulations, Barnett et al.~\cite{Barnett2011a} analyzed the spinor gas quench experiments \cite{Sadler2006a,Vengalattore2008a,Vengalattore2010a.PhysRevA.81.053612,Guzman2011a}, showing that the system did not thermalize at appreciable time scales but rather reached a quasi-steady regime that evolved anomalously slowly in time. 
The long-time evolution after the short-time instabilities and growth of magnetization \cite{Lamacraft2007.PhysRevLett.98.160404,Saito2007a.PhysRevA.75.013621,Saito2007b.PhysRevA.76.043613,Damski2007a.PhysRevLett.99.130402,Uhlmann2007a.PhysRevLett.99.120407,Cherng2008a.PhysRevLett.100.180404,Baraban2008a.PhysRevA.78.033609,Klempt2009a.PhysRevLett.103.195302,Sau2009a.PhysRevA.80.023622} was characterized by a quasi-steady state with exponentially decaying spatial correlations. 
The time scale for Landau damping was estimated to be beyond the reach of the experiment. 
It was concluded that the system did not thermalize at appreciable time scales but rather reached a quasi-steady regime that evolves anomalously slowly in time and which they called prethermalized.

A slow long-time residual growth of the spin domains created in the quenches across the ferromagnetic phase transition was reported in \cite{Guzman2011a} and conjectured to signal coarsening dynamics.
Coarsening as self-similar rescaling in time and space has been discussed  for many different systems within the theory of phase-ordering kinetics \cite{Bray1994a}, including spinor condensates \cite{Lamacraft2007.PhysRevLett.98.160404,Mukerjee2007a.PhysRevB.76.104519,Keekwon2011a.JStatMTE.P03013,Williamson2016a.PhysRevLett.116.025301}.  
These analyses typically assume dissipative dynamics, with, in some cases, additional `reversible' (non-dissipative) coupling between the modes \cite{Keekwon2011a.JStatMTE.P03013}.

\begin{figure}[t]
\flushright
\includegraphics[width=0.95\columnwidth]{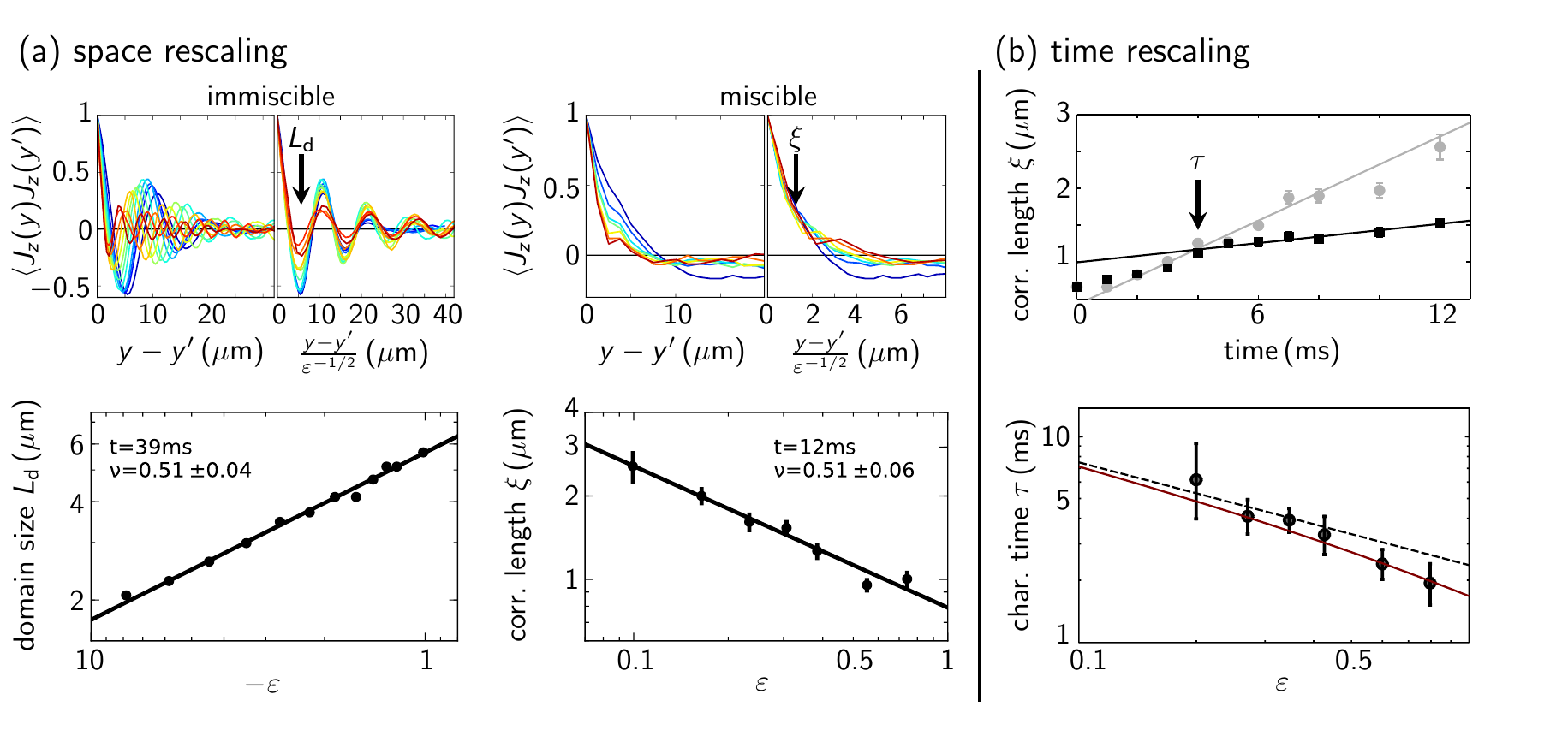}\ \ \vspace*{-3ex}
  \caption{Spatial and temporal scaling of the spin-spin correlations measured in a quasi 1D pseudo-spin $1/2$ gas \protect\cite{Nicklas:2015gwa} after a quench close to a quantum critical point.
  The quench is performed by suddenly changing the Rabi coupling $\Omega$ between the two components of the gas from a large value to one close to the critical value $\crit{\Omega}$ where a transition from a paramagnetic to a ferromagnetic phase is triggered.
(a) 
Spatial correlations after quenches to different relative proximities $\varepsilon=\Omega/\crit{\Omega}-1$ from the critical point (color-coded), on the ferromagnetic (left panels) and paramagnetic sides side (right) at the times indicated in the graphs. 
Top row: 
The correlation functions, under a rescaling $y\to y\varepsilon^{\nu}$ of the distance dependence with mean-field exponent $\nu=1/2$, fall on a universal curve.
Bottom row: 
$\varepsilon$-dependence of the characteristic length scales deduced from the correlation functions.
The straight lines reveal values for the critical exponent $\nu$ on either side of the transition.
(b) 
Temporal scaling of the spin-spin correlations.
After the quench to the miscible side of the transition, the correlation length $\xi$, measuring the exponential decay seen in the top right panel in (a), for any $\varepsilon\ge0$, shows first a linear rise and later settles in an oscillatory manner to a finite saturated value, see \cite{Nicklas:2015gwa} for more details.
The characteristic time $\tau$ for different $\varepsilon$ is obtained as the intersection point of the linear fits to the initial rise of $\xi(t)$ (grey symbols for $\varepsilon=0.1$) and to the behaviour after $\xi$ deviates from this rise.
The procedure of determining the intersection is exemplarily shown in the upper panel for $\varepsilon=0.23$. 
In the lower panel we compare the extracted $\tau(\varepsilon)$ to a mean-field scaling with $\nu z=1/2$ (dashed line) and to  the Bogoliubov prediction $\tau\sim1/\Delta$, with gap $\Delta(\varepsilon)=\crit\Omega\sqrt{\varepsilon(\varepsilon+1)}$, also applicable at larger $\varepsilon\gtrsim1$ (solid line). 
}
  \label{fig:spacetimescaling}
\end{figure}

In the experiments of Nicklas et al.~\cite{Nicklas2011a,Nicklas:2015gwa}, performing quenches to either side of a quantum critical point in a quasi 1D pseudo-spin 1/2 gas, scaling has been revealed of both, characteristic time and spatial scales with respect to the parametric distance of the final Hamiltonian to criticality, see \Fig{spacetimescaling}. 
Theoretical studies of these experiments, like of similar quench protocols in transverse-field Ising chains \cite{Calabrese2011a.PhysRevLett.106.227203,Calabrese2012a,Calabrese2012b}, indicated that the intermediate-time critical scaling properties of correlation lengths can be understood in terms of a GGE for a prethermalized state \cite{Karl2016a}.
For previous discussions of prethermalization and/or the approach of a GGE in quenches near criticality see also \cite{Sciolla2013a.PhysRevB.88.201110,Chandran2013a.PhysRevB.88.024306,Smacchia2015a.PhysRevB.91.205136,Chiocchetta2015a.PhysRevB.91.220302,Maraga2015a.PhysRevE.92.042151,Maraga2016a.160201763M} for isolated systems, or \cite{Gagel2014a.PhysRevLett.113.220401} for open systems.
Starting with a ground state far away from the critical point, the quench close to criticality maps this state onto new, nearly number-conserved quasiparticle degrees of freedom.
The resulting occupation numbers of these quasiparticles define the Lagrange parameters of the GGE.
It is nevertheless found that, the closer the quench of the pseudo-spin gas is tuned to the critical point, the better the correlations are described by a single effective temperature parameter \cite{Karl2016a}.
See~\cite{Calabrese2006a.PhysRevLett.96.136801,Calabrese2007a,Calabrese:2016xau} for earlier studies in conformal field theories.

In the light of fast experimental progress, the verification of  non-thermal fixed points as well as universal scaling dynamics as a more general kind of prethermalization in near-integrable quantum systems is at the horizon. 
This will include demonstrations of the universal coarsening of defects, strong non-linear excitations and turbulent dynamics.

\subsection{Impact of integrability on transport dynamics}
Integrability does not only affect the thermalization of cold atomic systems, but also has important effects on their transport dynamics. 
This was observed in an experiment by Ronzheimer et al.~\cite{Ronzheimer2013a.PRL110.205301} in which the expansion of initially localized atoms in homogeneous 1D and 2D optical lattices was studied. 
${}^{39}$K atoms were prepared in the combined potential of a 3D optical lattice and an additional harmonic confinement. 
Tunable inter\-actions allowed the realization of both the hard-core boson limit and bosons with finite interactions. 
The harmonic confinement was then decreased in one or two directions so that the atoms could expand in a 1D or 2D optical lattice potential. 
It was observed that the fastest ballistic expansion of hard-core bosons proceeds in all inte\-grable limits of the system, where the many constants of motion inhibit diffusive scattering. 
Deviations from these limits significantly suppressed the expansion and led to signatures of diffusive dynamics. 
Interestingly, for finite interactions, deviations from the ballistic expansion of hard-core bosons only occurred because the gas was subject to an interaction quench, implying the higher energies the weaker the interactions are  \cite{Sorg2014a}.

\section{Conclusions and outlook}
\label{sec:summary}

Advances in preparing and controlling low-dimensional ultracold gases have lead to an enormous, renewed interest in integrable and near-integrable model systems as prototypes for studying dynamics, relaxation, and thermalization in quantum physics. 
Insight from  experiments has lead to rapid progress in linking the microscopic quantum dynamics of atoms and molecules with the macroscopic properties of matter providing a unique connection between quantum many-body physics and statistical physics.
In particular they provide a unique window onto the emergence of (classical) statistical ensembles in the evolution of isolated many-body quantum systems and thus the transition from quantum physics at the micro-scale to our classical world.

The experimental progress in this field continues at a remarkable speed and will allow new insights into previously unattainable phenomena.
Examples include the ability to control the interaction strength~\cite{Timmermans1999b}, to observe and control quantum many-body systems locally at the single-atom level~\cite{Buecker2009a,Bakr2009a,Sherson2010a,Serwane2011a,Kaminishi2014a.1410.5576K}, the use of optimal control schemes to manipulate their
(global) external degrees of freedom \cite{Buecker2011a,Buecker2013a,vanFrank2014a,vanFrank2016a}. 
 Moreover, the extension to gases with fermionic statistics~\cite{Nascimbene2010a}, gauge fields~\cite{Goldman2014a}, spin-orbit coupling~\cite{Galitski2013a} or long-range interactions~\cite{Richerme2014a,Yan2013a,Kadau2016a} should bear many new interesting aspects. 
 
With these capabilites, ultracold atomic gases  offer themselves as a `quantum simulator' for universal dynamics of systems which are difficult to access directly~\cite{Bloch2012a.Simulation}. 
Beyond the immediate implications for simple low-energy degenerate quantum gases, phenomena such as topological configurations in solids, in soft matter, the dynamics of the quark-gluon plasma created in heavy-ion collisions, dynamics of the big bang \cite{Hung1213a.Science341.1213,Schmiedmayer:2013xsa}, or the reheating of the post-inflationary universe come into sight.

For the near future, we therefore expect a plethora of novel experimental and theoretical insights that will have profound implications for our understanding of the emergence of thermal and classical properties in isolated quantum many-body systems. 
This will lead to a truly universal framework for non-equilibrium dynamics, the study of which is an ongoing endeavor.

\section*{Acknowledgments}
The authors thank J.~Berges, F.~Brock, H.~Cakir, I.~Chantesana, S.~Czischek, E.~G.~Dalla Torre, M.~Davis, E.~Demler, J.~Eisert, S.~Erne, F.~Essler, C.~Ewerz, R.~Geiger, M.~Gring, A.~Johnson, M.~Karl, V.~Kasper, M.~Kastner, K.~Kheruntsyan, M.~Kronenwett, M.~Kuhnert, P.~Kunkel, D.~Linnemann, S.~Mathey, I.~Mazets, L.~McLerran, W.~Muessel, E.~Nicklas, B.~Nowak, M.~Oberthaler, J.~Pawlowski, A.~Pi{\~n}eiro Orioli, M.~Pr\"ufer, B.~Rauer, H.~Salman, A.~Samberg, C.~Scheppach, C.~Schmied, J.~Schole, T.~Schweigler, D.~Sexty, H.~Strobel, T.~Wright, and J.~Zill for discussions and collaboration on the topics described here. 
They acknowledge funding by the EU (SIQS, ERC advanced grant QuantumRelax, FET-Proactive grant AQuS, Project No.~640800), by the Austrian Science Fund (FWF), and by the National Science Foundation under Grant No. NSF PHY11-25915. T.L. acknowledges support by the Alexander von Humboldt Foundation through a Feodor Lynen Fellowship. 

\bibliography{Bibliography/Master}
\bibliographystyle{Bibliography/JHEP_notitle}

\providecommand{\href}[2]{#2}\begingroup\raggedright\endgroup

\end{document}